%% file: MMS-reviewFinal.tex
\documentclass[11pt,twoside]{article}
\usepackage{float,latexsym,graphicx,shortvrb,fancyhdr}
 \usepackage{amsmath,amsfonts,bm,mathtools,caption,booktabs}
\usepackage[a4paper]{geometry}
\usepackage{caption}
\usepackage{subcaption}
\input{command}
\renewcommand\bn{{\bf n}}

\def\la{\langle}
\def\ra{\rangle}

\def\G{G\^{a}teaux }

\renewcommand\PP{{\Pi}}
\newcommand\TT{{T}}
\newcommand\DD{{D}}
\newcommand\FF{{F}}
\renewcommand\xx{{\bf x}}

\renewcommand\Gu{{\Gamma_\chi}}

\newcommand\bh{{\bf h}}
\renewcommand\bff{{\bf f}}

\renewcommand\bvsig{{\vsig}}
\newcommand\WW{{W}}
\newcommand\VV{{V}}
\newcommand\UU{{U}}

\newcommand\bGd{{\bf G}_{\delta_k}}
\newcommand\bFd{{\bf F}_{\delta_k}}

\renewcommand\bw{{\beps}}

\newcommand\Z{Z\u{a}linescu }
\newcommand\dbchi{{\dot{\bchi}}}

\makeatletter
\renewcommand{\maketitle}{\bgroup\setlength{\parindent}{0pt}
\begin{flushleft}
  \textbf{\@title}

  \@author
\end{flushleft}\egroup
}
\makeatother

\begin{document}

$\;$\\
\begin{center}
{\Large \textbf{
Canonical Duality-Triality Theory:\\
 Bridge Between   Nonconvex Analysis/Mechanics and  Global Optimization in Complex Systems\\}}
\end{center}

\vspace{2em}
\noindent{{\bf\large David Yang Gao, Ning Ruan}\\
\em Faculty  of Sciences and Technology,
Federation University Australia, Mt Helen, VIC 3353, Australia.} \\

\noindent{{\bf\large Vittorio Latorre}\\
\em Department of Computer, Control and Management Engineering, �Sapienza�
University of Rome, Rome, Italy }\\

{\em Dedicated to Professor Gilbert Strang on the Occasion of His 80th Birthday}

\vspace{2em}
\noindent{\large\textbf{Abstract}}\\

Canonical duality-triality is a breakthrough
 methodological theory, which can be used not only for modeling complex systems within
a unified framework, but also for solving a wide  class of challenging problems from  real-world applications.
This paper presents a brief review on this theory, its philosophical origin, physics foundation, and
  mathematical statements in both finite and infinite dimensional spaces. 
Particular emphasis is placed on its
      role   for   bridging the gap between nonconvex analysis/mechanics and global optimization.
Special attentions are  paid on   unified
understanding the fundamental difficulties in large  deformation mechanics,  bifurcation/chaos in nonlinear science,
 and the  NP-hard problems in global optimization,
as well as the theorems, methods, and algorithms  for solving these challenging problems.
Misunderstandings and confusion on some basic concepts, such as objectivity, nonlinearity,
 Lagrangian, and generalized convexities are discussed and classified.
Breakthrough from recent  challenges  and conceptual mistakes  by
 M. Voisei, C.  Z\u{a}linescu and his co-worker  are addressed.
Some open problems and future works in global optimization and nonconvex mechanics   are proposed.\\

\noindent{\bf Keywords}: Duality, complementarity, triality,  mathematical modeling, large deformation,
nonlinear PDEs, NP-hard problems, nonconvex analysis, global optimization.

\vspace{.5em}

\section{Introduction}
Duality is one of the oldest and most beautiful ideas in human knowledge. 
It has a simple origin from the  oriental philosophy of
{\em yin-yang principle}  tracing back   
  5000 years ago.
According to {\em I Ching}\footnote{Also known as the {\em Book  of Changes,  Zhouyi} and {\em Yijing,}
is the world�s oldest and most sophisticated system of wisdom
divination, the fundamental source of most of the east�s philosophy, medicine and spirituality.
Traditionally it was believed that the principles of the I Ching originated with the mythical King Fu Xi
during the 3rd and 2nd millennia BCE.},
 the fundamental law  of the nature is the {\em Dao},   
 the duality of one yin and one yang, which  gives two    opposite  or complementary  points of view of looking at the same object.  %
In quantum mechanics,  the wave-particle duality is a typical example to fully  describe the behavior of   quantum-scale objects.
Mathematically,   duality represents certain translation of
concepts, theorems or mathematical structures  in a one-to-one fashion, i.e., if the dual of A is B, then the dual of B is A (cf. \cite{atiyah,cla-dav,morr}).
This one-to-one complementary  relation is called the {\em canonical duality}.
It is emphasized recently by  Sir Michael Atiyah that  duality in mathematics is not a theorem, but a ``principle"  \cite{atiyah}.
Therefore,  any duality gap is not allowed. This fact is well-known in mathematics and  physics, but not in   optimization
due to the existing gap between these  fields.
To bridge this gap,   a canonical duality-triality theory
has been developed originally from nonconvex   mechanics  \cite{gao-dual00} with 
  extensive applications 
in 
engineering,  mathematics, and  sciences,  especially in the multidisciplinary fields of  nonconvex mechanics
and  global optimization
\cite{gao-amma03,gao-cace09,gao-sherali-amma}. 

\subsection{Nonconvex analysis/mechanics and difficulties}

Mathematical theory of duality for convex problems has been well-established.
In linear elasticity,  it is well-known that each potential energy principle   is associated
with a unique  complementary energy principle through Legendre transformation.
This one-to-one  duality is guaranteed
 by  convexity of the stored  energy. 
 The well-known Helinger-Reissner principle is actually a special Lagrangian saddle min-max duality theory in convex analysis,
 which lays a foundation for mixed/hybrid finite element methods with successful applications in structural limit analysis \cite{gao-88a,gao-88b}.
 However, the one-to-one duality is broken in nonconvex systems.
 In large deformation theory, the stored energy  is generally nonconvex and its Legendre conjugate can't be uniquely determined.
 It turns out that   the existence of a pure stress-based
complementary-dual energy principle (no duality gap) was a well-known open problem over a half century
and subjected  to extensive discussions by many leading experts including Levison \cite{levi65}, Koiter \cite{koit},
Oden and Reddy \cite{oden-redd},  Ogden \cite{ogden75},
 Lee and Shield \cite{lee-shie80}, Stumpf \cite{stum}, etc.

 Nonconvex   phenomena arise naturally in large classes of engineering applications.
Many real-life problems in modern mechanics and complex systems require consideration of nonconvex
 effects  
 for their accurate modelling.
For example,  in  modelling of hysteresis, phase transitions, shape-memory alloys, and  super-conducting materials,
the free energy functions are usually nonconvex due to certain internal variables 
\cite{gao-anti,gao-ogden-qjmam,gao-ogden-zamp}. In large deformation analysis, thin-walled structure can
buckle even before the stress  reaches  its  elastic limit \cite{gao-sam94,gao-ejm95,gao-strang89b}.
Mathematically speaking,  many fundamentally difficult problems in engineering and the sciences are mainly due to the 
  nonconvexity of their modelling.
  In static systems, the nonconvexity usually leads to multi-solutions in the related governing equations.
Each of these solutions represents certain possible phase   or buckled state in large deformed solids.
 These local solutions are very sensitive to the internal parameters and external force.
 In dynamical systems, the so-called chaotic behavior is mainly due to nonconvexity of the objective functions
 \cite{gao-thes02}.
Numerical methods (such as FEM, FDM, etc) for solving nonconvex minimal potential variational problems usually end up with nonconvex
optimization problems \cite{gao-jem,gao-jogo00,gao-yu,ionita,santos-gao}. Due to the lack of global optimality criteria, finding  global optimal solutions is
fundamentally difficult, or even impossible by traditional numerical methods and optimization techniques.
For example, it was discovered  by Gao and Ogden \cite{ gao-ogden-qjmam,gao-ogden-zamp}
 that for certain given external loads, both the  global and local minimizers are  nonsmooth and
 cannot be determined by  any Newton-type numerical methods.
 In fact, many nonconvex problems are considered as NP-hard (Non-deterministic Polynomial-time hard) in global
 optimization and computer science \cite{gao-cace09,gao-sherali-amma}.
Unfortunately, these well-known difficulties are not fully recognized
in computational mechanics due to the significant gap between engineering mechanics and global
optimization. Indeed, engineers and scientists are mistakenly attempting to use traditional finite element
methods and commercial software for solving nonconvex mechanics problems. In order to identify the
fundamental difficulty of the nonconvexity from the traditional definition of nonlinearity, the terminology
of {\em  Nonconvex  Mechanics } was formally proposed by   Gao, Ogden and Stavroulakis in 1999   \cite{gao-ogden-stav}.
The {\em Handbook of Nonconvex Analysis}  by Gao and Motreanu  \cite{gao-motr} presents recent advances in the field.

\subsection{Global optimization and challenges} 
In parallel with the nonconvex mechanics, global optimization
is a multi-disciplinary research field developed  mainly from 
 nonconvex/combinatorial optimization and computational science
during the last nineties.
In general, the global optimization problem is formulated in terms of finding
 the absolutely best set of solutions for the   following   constrained optimization problem
\eb
\min f(x) , \;\; \mbox{ s.t. } \; h_i(x) = 0, \;\;  g_j(x) \le 0  \;\; \forall i \in I_m , \;\; j  \in I_p, \label{eq-go}
\ee
where   $f(x)$ is the so-called ``objective function"\footnote{This terminology is used mainly  in English literature.
The function $f(x)$ is called  the target function in Chinese and Japanese literatures,  the goal function in Russian and German literatures.},
$h_i(x)$ and $ g_j(x)$  are constraint functions,
 $I_m= \{1, \dots, m\} $ and $I_p=\{1, \dots, p\} $ are index sets.
It must be emphasized  that,  different from the the basic concept of {\em objectivity} in continuum physics,
the objective function extensively used in mathematical optimization is allowed to be any arbitrarily given function,
 even the linear function.
 Clearly, this mathematical model is artificial. Although it enables   one to ``model" a very wide range of problems,
  it comes at a price: even very special kinds of nonconvex/discrete optimization problems are considered to be NP-hard.
This dilemma   is due to the gap between mathematical optimization and  mathematical physics.
In science, the concept of objectivity is often attributed with the property of scientific measurements that can be measured independently of the observer.
Therefore, a  function in mathematical physics is called   objective  only if it depends on certain measure of its
variables (see Definition 6.1.2, \cite{gao-dual00} and the next section).
Generally speaking, a useful mathematical  model must   obey  certain fundamental law of nature.
Without detailed  information on these arbitrarily given functions,
it is impossible to have a general theory for finding global extrema of the general nonconvex problem (\ref{eq-go}). This could be the reason why there was no
breakthrough  in nonlinear programming during the past 60 years.

 In addition to the  nonconvexity,
  many  global optimization problems   in engineering design and operations research  explicitly require  integer or binary decision variables.
%
For example, in topology optimization of engineering structures, the design variable of material density
$\rho (\xx) = \{ 0, 1\} $ is a discrete selection field, i.e. by selection it has to take the value, 1,
and by de-selection it has to take the value, 0
 (see  \cite{bendsoe-sigmund}).
By the fact that the deformation variable  is a continuous field, which should be determined
  in  each iteration for  topological structure, therefore,
  the finite element method for solving topology optimization problems ends up with  a coupled
  mixed integer nonlinear programming problem.
   Discrete problems are frequently encountered in modeling real world systems for a wide spectrum of applications in
 decision science, management optimization, industrial and systems engineering.
 Imposing such integer constraints on the variables makes the global optimization problems much more difficult to solve.
  It is well-known in computational science and global optimization that even the most simple
 quadratic minimization problem with boolean constraint
  \eb
  \min
  \left\{\half \xx^T \bQ \xx - \xx^T \bff | \;\; \xx \in \{ 0, 1\}^n \right\} \label{eq-qip}
  \ee
  is considered to be NP-hard (Non-deterministic Polynomial-time hard) \cite{gao-ruan-jogo10}. Indeed, this integer minimization problem has
  $2^n$ local solutions. Due to the lack of global optimality criterion,
  traditional direct approaches, such as the popular branch and bound methods,
  can only handle very small size  problems.
  Actually, it was proved by Pardalos and Vavasis  \cite{pardalos91, vavasis90} that
  instead of the integer constraint,
  the continuous quadratic minimization with box constraints $  \xx \in [ 0, 1]^n $
  is NP-hard as long as   the matrix $\bQ $ has one negative eigenvalue.

 During the last 20 years, the field of global optimization has been developed dramatically to
  across almost every   branch of  sciences, engineering,  and complex systems   \cite{floudas,flod-pard,panos}.
  By the fact that the mathematical model  (\ref{eq-go}) is too general to have a mathematical theory for identifying global extrema,
  the main task in global optimization is to study  algorithmic methods for numerically  solving the
  optimal solutions.
  These methods  can be  categorized into two main  groups:    deterministic and stochastic.
  {\em Stochastic methods}  are based on an element of random choice. Because of this, one has to sacrifice the
possibility of an absolute guarantee of success within a finite amount of computation.
{\em  Deterministic methods,} such as the cutting plane,   branch and bound methods,  can find global optimal solutions, but
   not in polynomial time. Therefore, this type of methods can be used only for solving very small-sized problems.
   Indeed,  global optimization  problems with 200 variables
are referred to as ``medium scale",  problems with 1,000 variables as ``large scale", and the so-called
``extra-large scale" is only around 4,000 variables \cite{bura}.
In topology optimization, the variables could be easily 100 times more than this extra-large scale in global optimization.
 Therefore, to develop a unified deterministic theory   for efficiently  solving general global optimization problems
 is fundamentally important, not only in mathematical  optimization, but also in general nonconvex analysis and mechanics.




\section{Canonical Duality-Triality Theory}
The canonical duality-triality  theory   comprises mainly three parts:

   i) a {\em canonical dual transformation}, ii) a
{\em complementary-dual principle,} and iii) a {\em triality theory. }

 The canonical dual transformation is a versatile
methodology  which can be used to model complex systems within a unified framework
and to formulate perfect dual problems without a duality gap.
The complementary-dual principle presents a unified analytic solution form for general problems in continuous and
discrete systems. The triality theory reveals an intrinsic duality pattern in multi-scale  systems, which can be used
to identify both global and local extrema, and to develop
deterministic  algorithms for effectively solving a wide class of
nonconvex/nonsmooth/discrete optimization/variational problems.

\subsection{ General modeling and objectivity}

A useful methodological theory should  have solid foundations not only in physics, but also in mathematics,
even in philosophy and aesthetics.
The  canonical duality theory was developed from   Gao and Strang's original work for solving the following general   nonconvex/nonsmooth variational problem  \cite{gao-strang89}:
\eb
\min \{ \PP (\bchi) = W(D\bchi) - F(\bchi) \; | \; \bchi \in \calX_c   \}, \label{eq-gs}
\ee
where $F(\bchi)$ is the  external energy, which must be linear on its domain $\calX_a$;
the linear operator  $D:\calX_a \rightarrow \calW_a$ assigns each configuration $\bchi$ to an internal variable
 $\bw = D \bchi$ and,
 correspondingly,  $\WW:\calW_a \rightarrow \real$ is called the internal (or stored) energy.
 The feasible set $\calX_c = \{ \bchi \in \calX_a| \;\; D\bchi \in \calW_a\} $ is the  {\em kinetically  admissible space}.

 By Riesz representation theorem,  the external energy can be written as
  $F(\bchi) = \la \bchi, \barbchi^* \ra$, where $\barbchi^* \in \calX^*$ is a given input (or source).
  The bilinear form $\la \bchi, \bchi^* \ra :\calX \times \calX^* \rightarrow \real$ puts $\calX $ and $\calX^*$ in duality.
 Therefore, the  variation (or \G  derivative) of $F(\bchi)$ leads to  the
{\em action-reaction duality}: $\barbchi^* =   \partial  F(\bchi) $.
Dually,  the internal energy  must be an {\em  objective function} on its domain $\calW_a$
 such that the   intrinsic physical behavior  of the system can be described by the {\em constitutive duality}:
    $\bsig = \partial  W(\bw)$.

 Objectivity is  a basic concept  in mathematical modeling \cite{ciarlet,holz,marsd-hugh,ogden}, but is still
 subjected  to seriously study in  continuum physics  \cite{liu,murd,murd05}.
 The mathematical definition was given in Gao's book (Definition 6.1.2 \cite{gao-dual00}).
 \begin{definition}[Objectivity and Isotropy]
 Let
 $ {\cal R} $  be
 a proper orthogonal group, i.e. $\bR \in {\cal R} $ if and only if
 $  \bR^T = \bR^{-1} , \; \det \bR = 1$.
 A set $\calW_a   $ is said to be objective if
 \[
 \bR \bw \in \calW_a \;\; \forall \bw \in \calW_a, \; \forall \bR \in {\cal R}.
 \]
 A real-valued function $\WW:\calW_a \rightarrow \real$ is said to be objective if
 \eb
 \WW(\bR \bw ) = \WW(\bw) \;\; \forall \bw \in \calW_a, \; \forall \bR \in {\cal R}.
 \ee
  A set $\calW_a   $ is said to be isotropic if $
 \bw \bR \in \calW_a \;\; \forall \bw \in \calW_a, \; \forall \bR \in {\cal R}.$\\
 A real-valued function $\WW:\calW_a \rightarrow \real$ is said to be isotropic if
 \eb
 \WW(\bw \bR ) = \WW(\bw) \;\; \forall \bw \in \calW_a, \; \forall \bR \in {\cal R}.
 \ee
 \end{definition}

 Geometrically speaking, an objective function does not depend on the rotation, but only  on certain measure of its variable.
The isotropy means that the function $\WW(\beps)$  possesses a certain symmetry.
In continuum physics, the right Cauchy-Green  tensor\footnote{Tensor is a geometrical object
  which is defined as a multi-dimensional array satisfying a transformation law (see \cite{ogden}). A tensor must be independent of a particular choice of coordinate system (frame-indifference).
  But this  terminology has been misused in optimization literature, where,  any multi-dimensional array of data
  is called tensor (see \cite{bade}).} $\bC (\bF) = \bF^T \bF$ is an objective strain
measure, while the left Cauchy-Green tensor $\bc = \bF \bF^T$ is an isotropic strain measure.
In Euclidean space $\calW_a \subset \real^n$, the simplest objective function is the $\ell_2$-norm
$\|\bw\|$ in $\real^n$  as we have
$\|{\bf R} \bw \|^2 = \bw^T {\bf R}^T {\bf R} \bw = \|\bw\|^2 \;\;  \forall {\bf R} \in {\cal R}$.
In this case, the objectivity is equivalent to isotropy and,  in
Lagrangian mechanics, the kinetic energy   is required to  be isotropic \cite{land-lif}.

 Physically, an objective function doesn't depend on observers \cite{murd05}, which is essential for  any real-world mathematical
 modelling.  In continuum physics,
 objectivity
implies that the equilibrium condition of angular momentum (symmetry of the
Cauchy stress tensor $\bsig = \partial \WW(\beps)$, Section 6.1  \cite{gao-dual00}) holds.
It is emphasized by P.  Ciarlet that the objectivity is not an assumption, but an axiom \cite{ciarlet}.
  Indeed,  the objectivity is also known as the
  {\em axiom of material frame-invariance}, which   lays a foundation for the canonical duality theory.

 As an  objective function, the internal energy  $W(\bw)$ does not depends on each
  particular problem. Dually, the external energy $F(\bchi)$ can be called the {\em subjective function,}
   which depends on each given problem, such as the inputs, boundary conditions and  geometrical   constraints in $\calX_a$.
  Together, $\PP(\bchi) = W(D\bchi) - F(\bchi)$ is called the total potential energy and the minimal potential principle
  leads to the general optimization problem (\ref{eq-gs}).

  For dynamical problems, the liner operator $\DD = \{ \partial_t , \partial_x\}$ and
     $W(\DD\bchi) = \TT(\partial_t \bchi) - \VV(\partial_x \bchi)$, where   $\TT(\bv)$ is the kinetic energy   and $\VV(\be)$
     can be viewed as stored potential energy,   then
     \[
     \PP(\bchi) = \TT(\partial_t \bchi) - \VV(\partial_x \bchi) - \FF(\bchi)
     \]
     is the total action in dynamical systems.

  The necessary condition  $\delta \PP(\bchi) = 0$  for the solution of the minimization problem (\ref{eq-gs}) leads to
  a  general equilibrium equation:
  \eb
 A(\bchi) =  D^* \partial_{\beps}  W(D\bchi)   = \barbchi^* . \label{eq-geq}
  \ee
  This abstract form of equilibrium equation covers extensive  real-world applications ranging from
  traditional  mathematical physics, modern economics,   ecology, game theory, information technology, network optimization,
  operations research,    and much more
  \cite{gao-dual00,gao-sherali-amma,strang}.
  Particularly, if $\WW(\beps)$ is quadratic such that $\partial^2 \WW(\beps) = H$, then the operator $A:\calX_c \rightarrow \calX^*$
  is linear and can be written in the triality form: $A  = \DD^* H \DD$,
which appears extensively in mathematical physics,     optimization, and linear systems \cite{gao-dual00,oden-reddy,strang}.
Clearly, any convex quadratic function $\WW(\beps)$ is objective due to the
 Cholesky decomposition $A = \Lam^* \Lam \succeq 0 $.
  \begin{example}[Manufacturing/Production Systems]
  {\em
  In management science, the configuration variable is a vector $\bchi \in \real^n$, which could  represent  the products of a manufacture company.
  Its dual variable $\barbchi^* \in \real^n$ can be considered as market price (or demands). Therefore, the external energy
  $\FF(\bchi) = \la \bchi , \barbchi^*\ra = \bchi^T \barbchi^* $ in this example
is  the total income of the company.
The products are produced by workers $\beps \in \real^m$. Due to the cooperation, we have $\beps = \DD \bchi$ and
$\DD \in \real^{m\times n}$ is a matrix. Workers are paid by salary $\bsig = \partial \WW(\beps)$,
therefore, the internal energy $\WW(\beps)$ in this example is the cost, which should be an objective function.
Thus, $\PP(\bchi) = \WW(\DD \bchi) - \FF(\bchi)$ is the {\em total cost or target}  and the minimization problem $\min \PP(\bchi)$ leads to
the equilibrium equation
\[
\DD^T \partial_{\beps} \WW(\DD \bchi) = \barbchi^*,
\]
which is an algebraic equation in $\real^n$. The weak form of this equilibrium equation is 
$\la \bchi, D^T \bsig \ra = \la D \bchi ; \bsig \ra = \la \bchi, \barbchi^* \ra$, which is the well-known {\em D'Alembert's principle} or the {\em principle of virtual work} in Lagrangian mechanics. 
The cost function  $\WW(\beps)$ could be  convex for a very small company, but
usually nonconvex for big companies to allow some people having  the same salaries.
}
  \end{example}

  \begin{example}[Lagrange Mechanics] \label{exam-lag}
  {\em
  In analytical mechanics, the configuration $\bchi \in \calX_a \subset \calC^1[I; \real^{n}]$ is a continuous vector-valued
  function of time $t\in I \subset \real$. Its components $\{ \chi_i \} \; (i = 1, \dots, n) $ are  known as
  the {\em Lagrangian coordinates}\footnote{It is an unfortunate truth that
  many people don't know  the relation between  the Lagrangian  space $\real^n$ they  work in  and the Minkowski (physical) space   $\real^3\times \real$   they  live  in.}.  Its dual variable $\barbchi^*$ is the action vector function in $\real^n$, say $\bff(t)$.
  The external energy $\FF(\bchi) = \la \bchi, \barbchi^* \ra = \int_I \bchi(t) \cdot \bff(t) \dt$.
  While the internal energy $\WW(\DD\bchi)$ is the so-called   action:
  \[
  \WW(\DD \bchi) = \int_I L(\bchi, \dot{ \bchi} ) \dt , \;\; L=  \TT(\dot{ \bchi}  ) - \VV( \bchi)
  \]
  where $\TT $ is the kinetic energy density,  $\VV $ is the potential density, and $L= \TT - \VV$ is the standard
  {\em Lagrangian density}.
  In this case, the linear operator $\DD \bchi = \{ \partial_t, 1 \} \bchi = \{ \dot{\bchi}, \; \bchi\}$ is a vector-valued mapping.
     The kinetic energy $\TT $ must be an objective function of the velocity $\bv_k = \dot{\bx}_k(\bchi)$
     (or isotropic since $\bv_k$ is a vector) of each particle $\bx_k = \bx_k(\bchi) \in \real^3 \;\; \forall k \in I_m$, while the potential density $\VV$ depends on each problem.
   Together, $\PP(\bchi) = \WW(\DD \bchi) - \FF(\bchi)$   is called {\em total action}. Its  stationary condition leads to the {\em  Euler-Lagrange equation}:
   \eb
   \DD^* \partial\WW(\DD\bchi) =  - \partial_t \frac{\partial \TT( \dot{\bchi} )}{\partial \dot{\bchi}} -  \nabla  \VV( \bchi) = \bff. \label{eq-e-l}
   \ee
   For Newton mechanics, $\TT(\bv) = \half \sum_{k\in I_m}  m_k \|\bv_k\|^2 $ is quadratic, where $\|\bv_k\| $ represents the Euclidean norm
   (speed)    of the $k$-th particle in  $ \real^3$.
   For Einstein's special relativity theory, $\TT(\bv) = -m_0 c \sqrt{c^2 - \|\bv\|^2} $ is convex (see Chapter 2.1.2, \cite{gao-dual00}),
   where $m_0 > 0$ is the mass of a particle at rest, $c$ is the speed of light.
   Therefore, the total action $\PP(\bchi)$ is convex only if $\VV(\bchi)$ is linear. In this case, the solution of the
   Euler-Lagrange equation (\ref{eq-e-l})   minimizes the total action.
   The total action is nonconvex as long as the potential density $\VV(\bchi)$ is nonlinear.
   In this case, the system may have periodic solution if $\VV(\bchi)$ is convex and the well-known {\em least action principle is indeed a misnomer} (see Chapter 2, \cite{gao-dual00}).
The system may have chaotic solution  if the potential density  $\VV(\bchi)$ is nonconvex \cite{gao-na00,gao-amma03}.
 Unfortunately, these important facts  are  not well-realized in both classical mechanics and modern nonlinear dynamical systems. The recent  review article\cite{gao-bc15}
presents a  unified understanding 
bifurcation, chaos, and NP-hard problems in complex systems.

 } \end{example}

  In nonlinear analysis, the linear operator $\DD$ is a partial differential operator, say $\DD= \{ \partial_t, \; \partial_x\}$,
   and the abstract equilibrium equation (\ref{eq-geq})
  is a nonlinear partial differential equation.
For convex  $W(\bw)$,   the solution of this equilibrium equation is also a solution to the minimization problem (\ref{eq-gs}).
  However, for nonconvex $W(\bw)$, the solution of (\ref{eq-geq}) is only a  stationary point of $\PP(\bchi)$.
  In order to study   stability and regularity of the local solutions in nonconvex problems,
   many generalized definitions,    such as
quasi-, poly- and rank-one convexities have been introduced and subjected to extensively study for more than fifty years \cite{ball}.
But all these generalized convexities provide only local  extremality conditions, which lead to many   ``outstanding open problems"
  in nonlinear analysis \cite{ball}.
  However, by the canonical duality-triality theory, we can have clear understandings  on these challenges.

 \subsection{Canonical transformation and classification of nonlinearities}\label{sec-cdt}

According to the canonical duality,  the linear measure $\bw = D \bchi$ can't be used directly  for
studying constitutive law due to the objectivity.
Also,  the linear operator can't change the nonconvexity of $W(D\bchi)$.
Indeed, it is well-known that the deformation gradient $\bF = \nabla \bchi$ is not considered as a strain measure in nonlinear elasticity.
The most commonly used strain measure  is the  right Cauchy-Green strain tensor $\bC = \bF^T \bF$,
which is, clearly, an  objective function  since $\bC(\bF) = \bC(\bQ \bF)$.
According to P. Ciarlet (Theorem 4.2-1, \cite{ciarlet1}),
the stored energy $W(\bF)$ of a  hyperelastic material is objective if and only if there exists
a function $\tilde{W}$ such that $W(\bF) = \tilde{W}(\bC)$.
Based on this fact in continuum physics,
the canonical transformation is naturally introduced.

\begin{definition}[Canonical Function and Canonical Transformation]$\;$ \hfill

A real-valued function $\Phi:\calE_a \rightarrow \real$ is called canonical if the duality mapping
$ \partial \Phi: \calE_a \rightarrow \calE_a^*$ is one-to-one and onto.

For a given nonconvex function $W:\calW_a \rightarrow \real$, if there exists a geometrically admissible mapping
$\Lam:\calW_a \rightarrow \calE_a$ and a  canonical  function $\Phi:\calE_a \rightarrow \real$ such that
\eb
W(\bw) = \Phi( \Lam(\bw)),  \label{eq-ct}
\ee
then, the transformation (\ref{eq-ct}) is called the canonical transformation 
 and  $\bxi = \Lam(\bw)  $ is called the canonical  measure.
\end{definition}

By this definition, the  one-to-one duality relation
$\bxi^* = \partial \Phi(\bxi) : \calE_a \rightarrow
\calE^*_a$  implies that the  canonical function $\Phi(\bxi)$  is differentiable
 and its  conjugate function $\Phi^*:\calE^*_a \rightarrow \real$
 can be uniquely defined by the Legendre transformation \cite{gao-dual00}
\eb
\Phi^*(\bxi^*) = \{ \la \bxi ; \bxi^* \ra - \Phi(\bxi) | \; \bxi^* = \partial \Phi(\bxi) \} ,
\ee
where $\la \bxi ; \bxi^* \ra $ represents the bilinear form on $\calE$ and its dual space $\calE^*$.
In this case,  $\Phi:\calE_a \rightarrow \real$ is a canonical function if and only if
  the following canonical duality relations hold on $\calE_a \times \calE^*_a$:
 \eb
  \bxi^* = \partial \Phi(\bxi)  \;\; \Leftrightarrow \;\;   \bxi= \partial \Phi^*(\bxi^* ) \;\; \Leftrightarrow \;\;
  \Phi(\bxi) + \Phi^*(\bxi^*) = \la \bxi ; \bxi^* \ra. \label{eq-cdr}
  \ee

A canonical function  $\Phi(\bxi)$ can also be nonsmooth but should be convex such that its conjugate can be well-defined by Fenchel
transformation
\eb
\Phi^\sharp(\bxi^*) = \sup \{ \la \bxi ; \bxi^* \ra - \Phi(\bxi) | \;  \bxi \in \calE_a \} .
\ee
In this case, $\partial \Phi(\bxi) \subset \calE^*_a$ is understood as
the sub-differential  and  the canonical duality relations (\ref{eq-cdr}) should be written in the generalized form
\eb
  \bxi^* \in \partial \Phi(\bxi)  \;\; \Leftrightarrow \;\;   \bxi \in \partial \Phi^\sharp(\bxi^* ) \;\; \Leftrightarrow \;\;
  \Phi(\bxi) + \Phi^\sharp(\bxi^*) = \la \bxi ; \bxi^* \ra. \label{eq-cdrn}
  \ee
  This generalized canonical duality plays an important role in unified understanding Lagrangian duality and KKT theory
 for constrained optimization problems (see \cite{g-r-s,lato-gao-opl} and Section \ref{sec-const}).

In analysis, nonlinear PDEs are   classified as {  semilinear}, {   quasi-linear}, and {  fully nonlinear}
 three categories based on the degree of the nonlinearity  \cite{feng-etal}. A {\em semilinear PDE} is a differential
equation that is nonlinear in the unknown function but linear in all its
partial derivatives.    A {\em quasi-linear PDE} is one that is nonlinear in (at
least) one of the lower order derivatives but linear in the highest order derivative(s)
of the unknown function.  {\em Fully nonlinear PDEs } are referred
to as the class of nonlinear PDEs which are nonlinear in the highest order derivatives
of the unknown function.
However, this classification is not essential as we know that the main difficulty is nonconvexity, instead of nonlinearity
since these  nonlinear PDEs could be related to certain
convex variational problems, which can be solved easily by numerical methods.

The concepts of geometrical and physical nonlinearities are well-known in  continuum physics,
but not in  abstract analysis and  optimization.
This leads to many confusions.
Based on the canonical transformation, we can have the following classification.
\begin{definition}[Geometrical, Physical and Complete Nonlinearities] $\;$\hfill

The general problem (\ref{eq-gs}) is called geometrically nonlinear (resp. linear) if the geometrical operator $\Lam(\bw)$
is nonlinear (resp. linear);

The  problem (\ref{eq-gs}) is called physically nonlinear (resp. linear) if the constitutive relation $ \bxi^* = \partial \Phi(\bxi)$
is nonlinear (resp. linear);

The general problem (\ref{eq-gs}) is called completely nonlinear  if it is both geometrically  and physically nonlinear.
\end{definition}

According to this clarification,  the minimization problem (\ref{eq-gs}) is
 geometrically linear as long as  the stored energy $W(\bw)$ is convex. In this case, $\Lam(D\bchi) = D \bchi$ and
 $\Phi(\Lam(\bw) ) = W(\bw)$.
 Thus, a physically nonlinear  but geometrically linear  problem could be equivalent to a fully nonlinear PDE,
  which can be solved  easily by well-developed convex optimization techniques.
Therefore, the main difficulty in complex systems is the geometrical nonlinearity.
This is the reason why only this nonlinearity was  emphasized in the title of   Gao-Strang's paper \cite{gao-strang89}.
 The complete nonlinearity is also called fully nonlinearity in engineering mechanics.
Hope this new  classification will clear out this  confusion.
By the canonical  transformation, the  completely nonlinear minimization problem (\ref{eq-gs})
can be equivalently written in the following canonical form
\eb
(\calP): \;\;\; \min \{ \Pi(\bchi) = \Phi(\Lam( D \bchi)) - F(\bchi ) | \; \bchi \in \calX_c \}. \label{eq-canp}
\ee
In order to solving this nonconvex problem, we need to find its canonical dual form.

\subsection{Complementary-dual principle}
For geometrically linear problems, the stored energy $\WW(\bw)$ is convex and the complementary energy $\WW^*(\bsig)$ can be uniquely defined
on $\calW^*_a$ by Legendre transformation.
Therefore, by using  equality  $\WW(\bw) = \la \bw ; \bsig \ra - \WW^*(\bsig)$, the total potential $\PP(\bchi)$  can be equivalently written
in the classical Lagrangian form $L:\calX_a \times \calW_a^* \rightarrow \real$
\eb
L(\bchi, \bsig) = \la D \bchi ; \bsig \ra - W^*(\bsig) - F(\bchi) = \la \bchi , D^* \bsig - \barbchi^* \ra - W^*(\bsig),
\ee
where, $\bchi$ can be viewed as a Lagrange multiplier for the equilibrium equation $D^* \bsig = \barbchi^*$.
In linear elasticity, $L(\bchi, \bsig)$ is the well-known Hellinger-Reissner complementary energy.
 Let
 $\calS_c = \{ \bsig \in \calW^*_a | \; D^* \bsig = \barbchi^* \} $ be the so-called
 {\em statically admissible space}. Then the Lagrangian dual  of the general problem  (\ref{eq-gs})  is  given by
  \eb
 \max \{  \PP^*(\bsig) = - \WW^*(\bsig) | \; \bsig \in \calS_c \}, \label{eq-ld}
 \ee
and the following   Lagrangian min-max duality  is well-known:
\eb
\min_{\bchi \in \calX_c} \PP(\bchi) = \min_{\bchi \in \calX_a} \max_{\bsig \in \calW^*_a} L(\bchi, \bsig) =
  \max_{\bsig \in \calW^*_a} \min_{\bchi \in \calX_a} L(\bchi, \bsig) =
 \max_{\bsig \in \calS_c} \PP^*(\bsig)  .
\ee

In continuum mechanics, this one-to-one duality is called {\em complementary-dual variational principle} \cite{oden-reddy}.
In finite  elasticity,  the Lagrangian dual is also known as the {\em Levison-Zubov principle}.
However, this principle holds only for convex problems.
If the stored energy $\WW(\bw)$ is nonconvex, its complementary energy can't be determined uniquely by the Legendre transformation.
Although its  Fenchel conjugate $\WW^\sharp:\calW^*_a \rightarrow \real \cup \{ + \infty\}$ can be uniquely defined,
the Fenchel-Moreau  dual problem
\eb
\max \{ \PP^\sharp(\bsig) = - \WW^\sharp(\bsig) | \;\; \bsig \in \calS_c\}
\ee
is not considered as a complementary-dual problem due to Fenchel-Young inequality:
\eb
\min \{  \PP(\bchi)  | \; \bchi \in \calX_c \} \ge \max  \{ \PP^\sharp(\bsig)  | \;\; \bsig \in \calS_c\} ,
\ee
and $\theta = \min  \PP(\bchi) - \max \PP^\sharp (\bsig) \neq 0$ is the so-called {\em duality gap}.
This duality gap is intrinsic to all type of Lagrangian duality problems since  the nonconvexity of $\WW(D\bchi)$
can't be changed by any linear operator.
It turns out that the existence of a pure stress based complementary-dual principle  has been a well-known debet in finite elasticity
for more than forty years \cite{li-gupta}.

\begin{remark}[Lagrange Multiplier Law]
{\em
Strictly  speaking, the Lagrange multiplier method can be used mainly for equilibrium constraint
 in $\calS_c$  and  the Lagrange multiplier must be the solution to the primal problem (see Section 1.5.2 \cite{gao-dual00}).
 The equilibrium equation $\DD^* \bsig = \barbchi^*$
  must be an invariant under certain coordinates transformation,
say the law of angular momentum conservation,
 which is  guaranteed by  the objectivity of  the stored energy $\WW(\DD\bchi)$ in continuum mechanics (see Definition 6.1.2, \cite{gao-dual00}), or by the
 isotropy  of the kinetic  energy  $\TT(\dbchi)$ in Lagrangian mechanics \cite{land-lif}.
 Specifically,  the equilibrium equation for Newton's mechanics is an invariant under the Calilean transformation;
 while for Einstein's special relativity theory, the  equilibrium equation $\DD^* \bsig = \barbchi^*$ is an invariant under the
 Lorentz transformation.
 For linear equilibrium equation, the  quadratic
 $\WW(\beps)$ is naturally an objective function for convex systems.
    Unfortunately, since  the concept of the  objectivity is misused in mathematical optimization,
    the Lagrange multiplier method has been mistakenly used for solving general nonconvex problems, which produces
    many  different  duality gaps.
     }
     \end{remark}
In order to recover the duality gap in nonconvex problems, we use the canonical transformation $\WW(\DD\bchi) = \Phi(\Lam(\DD\chi))$ such that the nonconvex total potential $\Pi(\bchi)$ can be reformulated as
 the   total complementary energy $\Xi: \calX_a \times \calE^*_a \rightarrow \real$
\eb
\Xi(\bchi, \bxi^*) = \la \Lam (D\bchi) ; \bxi^* \ra - \Phi^*(\bxi^*) - F(\bchi),
\ee
which   was first introduced by Gao and Strang in 1989 \cite{gao-strang89}.
The stationary condition $\delta \Xi(\bchi, \bxi^*) = 0$ leads to the following canonical  equations:
\begin{eqnarray}
\Lam(D\bchi) = \partial \Phi^*(\bxi^*) , \label{eq-canc}\\
D^*\Lam_t(D\bchi) \bxi^* = \partial F(\bchi) ,\label{eq-cane}
\end{eqnarray}
where $\Lam_t(\bw) = \partial \Lam(\bw)$ is a generalized \G derivative of $\Lam(\bw)$.
By the canonical duality, (\ref{eq-canc}) is equivalent to
$\bxi^* = \partial_\bxi \Phi(\Lam(D \bchi))$.
Therefore, the canonical equilibrium equation  (\ref{eq-cane}) is the general equilibrium equation (\ref{eq-geq}).

By using the Gao-Strang complementary function, the canonical dual of $\Pi(\bchi)$ can be obtained  as
\eb
\Pi^d(\bxi^*) = \sta \{ \Xi(\bchi, \bxi^*) | \;\; \bchi \in \calX_a \} = \FF^\Lam (\bxi^*) - \Phi^*(\bxi^*),
\ee
where $\FF^\Lam(\bxi^*)$ is the $\Lam$-transformation defined by \cite{gao-jogo00}
\eb
\FF^\Lam(\bxi^*) = \sta \{ \la \Lam(D \bchi) ; \bxi^* \ra - \FF(\bchi) | \; \bchi \in \calX_a \}.
\ee
Clearly, the stationary condition in this $\Lam$-transformation is the canonical equilibrium equation (\ref{eq-cane}).
Let $\calS_c \subset \calE^*_a$ be a feasible set, on which  $\FF^\Lam(\bxi^*)$ is well-defined. Then we have the following
result.
\begin{thm}[Complementary-Dual Principle \cite{gao-mrc99,gao-mecc99}]
If $(\barbchi, \barbxi^*) \in \calX_a \times \calE^*_a  $ is a stationary point of $\Xi(\bchi, \bxi^*)$, then
$\barbchi$ is a stationary point of $\Pi(\bchi) $ on $\calX_c$, while  $\barbxi^*$ is a stationary point of $\Pi^d(\bxi^*)$ on $\calS_c$,
and
\eb
\Pi(\barbchi) = \Xi(\barbchi, \barbxi^*) = \Pi^d(\barbxi^*).
\ee
\end{thm}

This theorem shows that there is no duality gap between $\Pi(\bchi)$ and $\Pi^d(\bxi^*)$.
 In many real-world applications, the geometrical operator $\Lam(\bw)$ is usually quadratic such that
 the total complementary function   $\Xi(\bchi, \bxi^*)$ can be written as
 \eb
 \Xi(\bchi, \bxi^*) = \half \la \bchi , \bG(\bxi^*) \bchi \ra - \Phi^*(\bxi^*) - \la \bchi, \bF(\bxi^*)\ra  \label{eq-xiq}
 \ee
 where $\bG(\bxi^*) = \nabla^2_{\bchi}  \Xi(\bchi, \bxi^*)$ and $\bF(\bxi^*)$ depends on the linear terms in $\Lam(\DD\bchi)$ and the input $\barbchi^*$.
 The first term in  $\Xi(\bchi, \bxi^*)$
 \eb
 G_{ap}(\bchi, \bxi^*) =  \half \la \bchi , \bG(\bxi^*) \bchi \ra
 \ee
 is the so-called {\em complementary gap function} introduced by Gao and Strang in \cite{gao-strang89}.
 In this case, the canonical equilibrium equation $\nabla_\bchi \Xi(\bchi, \bxi^*) =  \bG(\bxi^*) \bchi - \bF(\bxi^*) = 0$ is linear in $\bchi$
 and the canonical dual $\Pi^d$ can be explicitly formulated as
 \eb
 \Pi^d(\bxi^*) = -  G^*_{ap}(\bxi^*)  - \Phi^*(\bxi^*),  \label{eq-cdg}
 \ee
 where
$G^*_{ap} (\bxi^*) =  \half \la  \bG^{-1} (\bxi^*) \bF(\bxi^*) ,  \bF(\bxi^*) \ra $
is called {\em pure complementary gap function}.
 Comparing this canonical dual with the Lagrangian dual $\PP^*(\bsig) = - \WW^*(\bsig)$ in (\ref{eq-ld}) we can find that
 in addition to replace $\WW^*$ by the canonical dual $\Phi^*$, the first term in $\Pi^d$ is
 identical to the Gao-Strang complementary gap function, which recovers the duality gap in Lagrangian duality theory
 and plays an important role in triality theory.

 \begin{thm}[Analytical Solution Form]
 If $\barbxi^* \in \calS_c$ is a stationary  point of $\Pi^d(\bxi^*)$, then
 \eb
 \barbchi = \bG^{-1} (\barbxi^*)  \bF(\barbxi^*) \label{eq-anas}
 \ee
 is a stationary point of $\Pi(\bchi)$ on $\calX_c$ and $\Pi(\barbchi) = \Pi^d(\barbxi^*)$.
 \end{thm}

This theorem shows that the  primal  solution  is analytically depends on its
canonical dual solution. Clearly,  the  canonical dual of a nonconvex primal problem  is also nonconvex and may
have multiple stationary points. By the canonical duality, each of these stationary solutions is corresponding to a primal solution
via (\ref{eq-anas}). Their extremality is governed by  Gao and Strang's  complementary gap function.

\subsection{Triality theory}\label{sec-tri}
In order to identify extremality of these stationary solutions,
we need to assume that the canonical function $\Phi:\calE_a \rightarrow \real$ is convex and let
\eb
\calS_c^+ = \{ \bxi^* \in \calS_c| \;\; \bG (\bxi^*) \succ 0 \} , \;\; \;\;\calS_c^- = \{ \bxi^* \in \calS_c | \;\; \bG (\bxi^*) \prec 0 \}
.
\ee
Clearly, for any given $\bchi \in \calX_a $ and $ \bchi \neq0 $, we have
\[
G_{ap}(\bchi, \bxi^*) >  0 \;\;    \Leftrightarrow \;\; \bxi^* \in \calS_c^+, \;\;
G_{ap}(\bchi, \bxi^*) < 0 \;\;   \Leftrightarrow \;\; \bxi^* \in \calS_c^-.
\]
\begin{thm}[Triality Theorem]\label{thm-tri}
Suppose $ \barbxi^* $ is a stationary point of $\Pi^d(\bxi^*)$ and
$\barbchi = \bG^{-1} (\barbxi^*) \barbxi^*$. If $\barbxi^* \in \calS_c^+$, we have
\eb
\Pi(\barbchi) = \min_{\bchi \in \calX_c} \Pi(\bchi)  \;\; \Leftrightarrow \;\; \max_{\bxi^* \in \calS_c^+} \Pi^d(\bxi^*) = \Pi^d(\barbxi^*);\label{eq-tris}
\ee
If $\barbxi^* \in \calS_c^-$, then on a neighborhood\footnote{The neighborhood $\calX_o$ of $\barbchi$  means that on which,
$\barbchi$ is the only stationary point.}   $\calX_o\times \calS_o \subset \calX_c\times \calS_c^-$ of $(\barbchi, \barbxi^*)$,
we have either
\eb
\Pi(\barbchi) = \max_{\bchi \in \calX_o} \Pi(\bchi)  \;\; \Leftrightarrow \;\; \max_{\bxi^* \in \calS_o} \Pi^d(\bxi^*) = \Pi^d(\barbxi^*) , \label{eq-trima}
\ee
or (only if $\dim \barbchi = \dim \barbxi^*$)
\eb
\Pi(\barbchi) = \min_{\bchi \in \calX_o} \Pi(\bchi)  \;\; \Leftrightarrow \;\; \min_{\bxi^* \in \calS_o} \Pi^d(\bxi^*) = \Pi^d(\barbxi^*)\label{eq-trimi}.
\ee
\end{thm}

The first statement (\ref{eq-tris}) is  called {\em canonical min-max duality}. Its weak form 
  was discovered by Gao and Strang in 1989 \cite{gao-strang89}. This duality
 can be used to identify global minimizer of the nonconvex problem (\ref{eq-gs}).
According this statement, the nonconvex problem (\ref{eq-gs}) is equivalent to
the following canonical dual problem, denoted by $(\calP^d)$:
\eb
(\calP^d): \;\;\; \max \{ \Pi^d(\bxi^*) | \; \bxi^* \in \calS_c^+  \}. \label{eq-cdmax}
\ee
This is a concave maximization problem which can be solved easily by well-developed convex analysis and optimization techniques.
The second statement (\ref{eq-trima}) is the {\em canonical double-max duality} and
  (\ref{eq-trimi}) is the  {\em canonical double-min duality}.
  These two statements can be used to identify the biggest local maximizer and local minimizer of the primal problem, respectively.

The  triality theory was first discovered by Gao 1996 in
 post-buckling analysis of a large deformed beam \cite{gao-amr97}. The generalization  to global optimization
 was made in 2000 \cite{gao-jogo00}.
It was realized in 2003 that the double-min duality (\ref{eq-trimi}) holds under certain additional condition \cite{gao-amma03,gao-opt03}.
Recently, it is proved that this  additional condition is simply $\dim \barbchi = \dim \barbxi^*$
to have the strong canonical double-min duality (\ref{eq-trimi}), otherwise,
this  double-min duality holds weakly in subspaces of $\calX_o\times \calS_o$
\cite{gao-wu-jimo,gao-wu-jogo,mora-gao-memo,mora-gao-naco}.
\begin{example}
  {\em

To  explain the theory, let us  consider a very simple nonconvex optimization in
$\real^n$:
\begin{eqnarray}
\min \left\{ \PP(\bx)=\half \alpha \left(\half\|\bx\|^2-\lam \right)^2-\bx^T \bff  \; \;\; \forall \bx \in \real^n \right\} , \label{eq-exam1}
\end{eqnarray}
where $\alp, \lam > 0$ are given parameters.
The criticality condition $\nabla P(\bx)=0$ leads to a nonlinear algebraic
equation system in $\real^n$
\begin{eqnarray}
\alpha (\half \|\bx\|^2-\lam)\bx =\bff.
\end{eqnarray}
Clearly, to solve this nonlinear algebraic equation directly   is difficult.
Also traditional convex optimization theory
 can not be used to identify global minimizer.  However, by the
canonical dual transformation, this problem can be solved completely and easily.
To do so, we let  $\xi=\Lam(\bx)=\half\|\bx\|^2  \in \real$, which is an objective measure. Then,
the nonconvex function $W(\bx) = \half \alpha(\half \| \bx \|^2 -\lam)^2$
can be written in canonical form
$\Phi(\xi) = \half \alpha (\xi - \lam)^2$.
Its Legendre conjugate is  given by
$\Phi^{\ast}(\vsig)=\half \alpha^{-1}\vsig^2 + \lam \vsig$, which is strictly convex.
Thus,
the total  complementary function for this nonconvex optimization
problem is
\begin{eqnarray}
\Xi(\bx,\vsig)= \half \|\bx\|^2   \vsig-\half
\alpha^{-1}\vsig^2 - \lam \vsig - \bx^T \bff.
\end{eqnarray}
For a fixed $\vsig \in \real$, the  criticality condition
$\nabla_{\bx} \Xi(\bx)=0$ leads to
\begin{eqnarray}\label{balance}
\vsig \bx-\bff=0.
\end{eqnarray}
For each  $\vsig \neq 0  $,
the  equation
(\ref{balance}) gives $\bx=\bff/\vsig$ in vector form. Substituting this into the
total complementary function $\Xi$,
the canonical dual function can be easily obtained as
\begin{eqnarray}
\Pi^d(\vsig)= \{\Xi(\bx,\vsig)| \nabla_{\bx} \Xi(\bx,\vsig)
=0\}
=  -\frac{\bff^T \bff}{2 \vsig}-\half \alpha^{-1} \vsig^2
-\lam \vsig, \;\;\; \forall \vsig\neq 0.
\end{eqnarray}
The critical point of this canonical function is obtained
by solving the following dual algebraic  equation
\begin{eqnarray}
2 (\alpha^{-1} \vsig+\lam)\vsig^2= \bff^T \bff. \label{eq-deuler}
\end{eqnarray}
For any given parameters $\alpha$, $\lam$ and the vector $\bff\in \real^n$,
this cubic algebraic equation has at most three real  roots
satisfying $\vsig_1 \ge 0 \ge \vsig_2\ge \vsig_3$,
and each of these roots leads to a critical point of the nonconvex function
$P(\bx)$, i.e., $\bx_i=\bff/\vsig_i$, $i=1,2,3$. By
the fact that $\vsig_ 1 \in \calS^+_c = \{ \vsig \in \real\; |\; \vsig > 0 \}$,
$\vsig_{2,3} \in \calS^-_c = \{ \vsig \in \real\; |\; \vsig < 0 \}$,
then Theorem \ref{thm-tri} tells us that $\bx_1$ is
a global minimizer of $\Pi(\bx)$, $\bx_3 $ is a local maximizer of $\Pi(\bx)$,
while $\bx_2$ is a local minimizer if $n=1$ (see Fig. 1).
If we choose    $n = 1, \;\; \alpha= 1$, $\lam=2$, and $f= \half$,
the primal function and canonical dual function
are shown in Fig. \ref{onedim} (a), where,  $x_1= 2.11491$ is global minimizer
of $\Pi(\bx)$, $\vsig_1=0.236417$ is global maximizer of $\Pi^d(\bvsig)$, and
$\Pi(x_1)=-1.02951=\Pi^d(\vsig_1)$ (see the two black dots).
Also it is easy to verify that $x_2 $ is a local minimizer, while $x_3$ is a local maximizer.
\begin{figure}
\includegraphics[scale=.55]{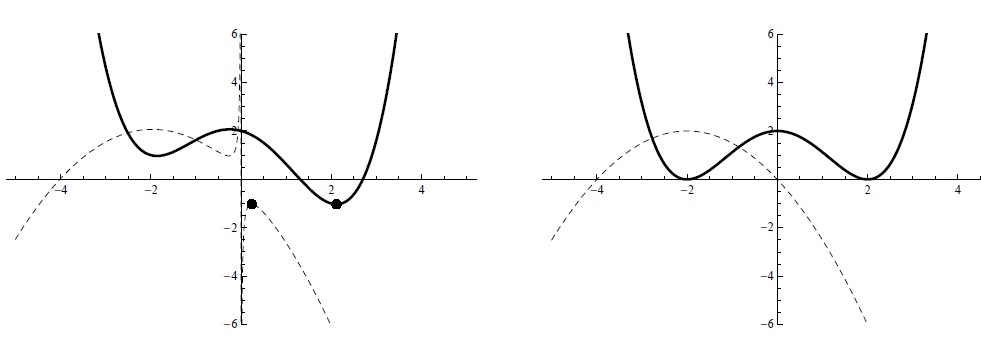}
 $  \mbox{ (a) $f = 0.5$ }$  \hspace{5cm} (b) $f = 0.\;\;\;\;\;\;$
\centering
\caption{ Graphs of   $\Pi(\bx)$  (solid) and $\Pi^d(\vsig)$ (dashed) }\label{onedim}
\end{figure}

If we let $\bff= 0$, the graph of $\Pi(\bx)$ is symmetric (i.e. the so-called double-well potential or the Mexican hat for $n=2$
\cite{gao-amma03})
 with  infinite number of global minimizers
 satisfying $\| \bx \|^2  = 2 \lam$.
In this case, the canonical dual $\Pi^d (\vsig) = - \half \alp^{-1} \vsig^2 - \lam \vsig$ is strictly concave
with only one critical point (local maximizer) $\vsig_3 = -  \alp \lam  \in \calS_c^-$ (for $\alp, \lam > 0$).
The corresponding solution $\bx_3 = \bff /\vsig_3 = 0$ is a local maximizer.
 By the canonical dual equation (\ref{eq-deuler})
we have $\vsig_1 = \vsig_2 = 0$ located on the boundary of $\calS^+_c$,
which  corresponding to the two global minimizers   $x_{1,2} = \pm \sqrt{2 \lam}$ for $n=1$, see Fig. 1 (b).
This is similar to the post-buckling of large deformed beam.
Due to symmetry $(f = 0)$, the nonconvex function $\Pi(\bx)$ has two possible buckled solutions  $\bx_{1,2} =  ( \pm \sqrt{2 \lam}, 0 )$
  with
the axial load $\lam = \half (b^2 - a^2)$.
While the local maximizer $\bx_3 = \{0, 0\}$ is corresponding to the unbuckled state.

This simple example shows a fundament issue in global optimization, i.e.,
the  optimal solutions of a nonconvex problem depends sensitively on the linear term (input or perturbation) $\bff$.
Geometrically speaking, the objective function $\WW(\DD \bx)$
in $\Pi(\bx)$ possesses certain symmetry.
If there is no linear term  (subjective function) in $\Pi(\bx)$,
 the nonconvex problem usually has  more than one
global minimizer due to the symmetry.
Traditional direct approaches and the popular SDP method are usually failed to deal with this situation.
By the canonical duality theory, we understand that in this case the canonical dual function $\Pi^d(\vsig)$  has no critical point in
 $\calS_c^+$.
Therefore, the  input  $\bff$  breaks
 the  symmetry so that $\Pi^d(\vsig)$ has a unique stationary point in $\calS_c^+$ which can be obtained easily.
This idea was originally from Gao's work (1996) on post-buckling analysis of large deformed beam \cite{gao-mrc96}, where the triality theorem was first proposed  \cite{gao-amr97}. The potential energy of this beam model is a double-well function, similar to this   example,
  without lateral force or imperfection, the beam could have two buckling states
(corresponding to two minimizers) and one un-buckled state (local maximizer).
Later on (2008) in the Gao and Ogden  work on analytical solutions in phase transformation \cite{gao-ogden-qjmam},
 they further discovered that the nonconvex system has no phase transition unless the  force distribution
  $f(x)$ vanished at
certain points.
 They also discovered that if force field $f(x) $ changes dramatically, all the Newton type direct approaches
failed  even to find any  local minimizer.
The linear perturbation method has been used successfully for solving global optimization problems \cite{chen-gao-oms,mora-gao-naco,r-g-j,wang-ea}.
}\end{example}

\section{Applications for  modeling of complex systems}
By the fact that the canonical duality is a fundamental law governing natural phenomena
and the objectivity is a basic condition for mathematical models,
the canonical duality-triality theory can be used for modeling real-world problems within a unified framework.

\subsection{Mixed integer nonlinear programming }
The most general and challenging problem in global optimization could be the mixed integer nonlinear program (MINP),
which is a minimization problem generally formulated as (see \cite{gros})
\eb
\min \{ f( \bx,\by) | \;\; g_i (\bx,\by) \le 0 \;\; \forall i \in I_m, \;\;  \bx \in  \real^n, \;\; \by \in {\mathbb{Z}}^p \} \label{eq-minp}
\ee
where ${\mathbb{Z}}^p$ is an integer set,
the ``objective function" $f(\bx,\by)$ and constraints $g_i(\bx,\by)$ for $i\in I_m$ are arbitrary functions \cite{burer-le}.
 Certainly, this  artificial model is virtually applicable to any problem in operations research, but it is impossible to develop
 a general theory and  powerful algorithm without detailed information given on these functions.
 As we know that the objectivity is a fundamental concept  in mathematical modeling.
 Unfortunately, this concept   has been mistakenly used with other functions,
such as target, cost, energy, and utility functions, etc\footnote{See $http://en.wikipedia.org/wiki/Mathematical_{-}optimization$}.

Based on the  Gao-Strang model (\ref{eq-gs}), we let  $\bchi = (\bx,\by), \; \bD\bchi = ( \bD_x \bx, \; \bD_y \by)$, and
$\barbchi^* = ( \bb, \bt )$. Then the general  MINP problem  (\ref{eq-minp}) can be remodeled  in the following form
\eb
\min \{ \PP(\bx,\by) = \WW(\bD_x \bx, \bD_y \by) -   \bx^T \bb  -   \by^T  \bt \;  | \;\;
(\bx,\by) \in \calX_c \times {\calY}_c, \;\; \bx \in  {\mathbb{Z}}^p \},
\ee
where the feasible sets are,  correspondingly,
\[
\calX_c = \{ \bx \in \calX_a \subset \real^n \; | \;\; \bD_x \bx \in \calU_a \},
\; \; \calY_c = \{ \by \in \calY_a \subset \real^p  | \; \bD_y \by \in \calV_a \}.
\]
In $\calX_a, \calY_a$, certain linear constraints are given, while in $\calU_a , \; \calV_a$, general nonlinear (constitutive) constraints
are prescribed such that the nonconvex (objective) function $\WW:\calU_a \times \calV_a \rightarrow \real$ can be written in
the canonical form
$\WW(\bD \bchi)  = \Phi_\chi(\bLam(\bchi))$
for certain geometrical operator $\bLam(\bchi)$.
 By the fact that any integer set ${\mathbb{Z}}^p$ is  equivalent   to a Boolean set \cite{ruan-gao-jogo14,wang-ea1},
 we simply let  ${\mathbb{Z}}^p=   \{ 0, 1 \}^p$.
 This constitutive constraint can be relaxed by the canonical transformation  \cite{gao-jimo07,gao-ruan-jogo10}
\eb
\beps =  \bLam_x(\bx) = \bx \circ (\bx - {\bf 1}) = \{ x_i^2 - x_i \}^p,
\ee
and the canonical function $\Phi_x(\beps) = \{ 0 \mbox{ if } \beps =  {\bf 0}, \; \infty \mbox{ otherwise} \}$.
Therefore, the canonical form for the MINP problem is
\eb
\min \{ \Pi(\bx,\by) = \Phi_\chi(\bLam(\bx,\by)) + \Phi_x(\bLam_x(\bx)) - \bx^T \bb  -   \by^T  \bt  \;  | \;
(\bx,\by) \in \calX_c \times \calY_c\}.
\ee
This canonical form covers many real-world applications, including  the so-called fixed cost problem \cite{g-r-s-fix}.
By the fact that the  canonical function $\Phi_x(\beps)$  is convex, semi-continuous,
the canonical duality relation should be replaced by
the sub-differential form $\bsig \in \partial \Phi_x(\beps) $, which is equivalent to
\eb
\bsig^T \beps =  0  \;\;\Leftrightarrow    \;\;  \beps  = {\bf 0} \;\; \forall \bsig \neq {\bf 0} .
\ee
Thus,  the integer constraint $\beps = \bLam_x(\bx) = \{x_i(x_i -1)\} = {\bf 0}$ can be relaxed by the
 canonical dual constraint $\bsig \neq {\bf 0}$ in continuous space.

The canonical duality-triality theory has be used successfully for solving mixed integer programming problems
\cite{chen-gao-ruan,gao-ruan-jogo10,g-r-s-fix}. Particularly, for  the quadratic integer programming problem (\ref{eq-qip}), i.e.
\[
\min \left\{ \PP(\bx) = \half \bx^T \bQ \bx - \bx^T \bff | \;\; \bx \in \{ 0, 1 \}^n \right\},
\]
the canonical dual is \cite{fang-gao-h-w06,gao-jimo07}
\eb
\max \left \{ \Pi^d(\bsig) = -\half ( \bff + \bsig)^T \bG^{-1}(\bsig)  (\bff + \bsig)   | \;\; \bsig \in \calS_c^+ \right\}
\ee
where $\bG (\bsig) = \bQ + 2 \Diag (\bsig) $. This is a concave maximization problem
over the convex set in continuous space
\[
\calS_c^+ = \{ \bsig \in \real^n | \; \bsig \neq {\bf 0} , \;\; \bG(\bsig) \succ 0 \},
\]
which can be solved easily if   $\calS_c^+ \neq \emptyset $.
Otherwise,  the integer programming  problem (\ref{eq-qip}) could be NP-hard, which is a conjecture proposed in \cite{gao-jimo07}.
In this case, a second  canonical dual problem has been proposed in \cite{gao-cace09,gao-watson-etal}
\eb
\min\left\{ \Pi^g(\bsig) = - \half \bsig^T \bQ^{-1} \bsig - \sum_{i=1}^n | f_i - \sig_i| \;\; | \; \bsig \in \real^n \right\}.
\ee
This is a unconstrained nonsmooth minimization problem, which can be solved by some deterministic methods, such as DIRECT method
\cite{gao-watson-etal}.

\begin{remark}[Subjective Function and NP-hard Problems] {\em
The subjective function $\FF(\bchi) = \la  \bchi , \barbchi^* \ra $ in the general model
$ \PP(\bchi) = \WW(\DD\bchi) - \FF(\bchi)$  plays an important role in
global optimization problems.
It was proved in \cite{gao-cace09} that for quadratic integer programming problem (\ref{eq-qip}), if the
source term $\bff$ is bigger enough, the solution is simply $\{ x_i\} = \{0  \mbox{ if }  f_i  < 0 , \; 1 \; \mbox{ if } f_i > 0\}$ (Theorem 8, \cite{gao-cace09}).
 If a system has no  input, by Newton's law, it  has either trivial solution or infinite number solutions.
For example, the  well-known max-cut problem
 \eb
 \max \left\{ \PP(\bx) = \frac{1}{4} \sum_{i,j = 1}^{n+1} \omega_{ij} (1 - x_i x_j) \; | \; x_i \in \{-1, 1\} \forall i = 1, \dots, n\right\}
 \ee
 is a special case of quadratic integer programming problem without the linear term. The integer condition
  is a physical (constitutive) constraint. Since there is no geometrical constraint, the graph is not fixed and any rigid motion is possible. Due to the symmetry $\omega_{ij} = \omega_{ji} > 0$,
  the global solution is not unique.
   The canonical dual feasible space $\calS^+_c$ in this example is empty and the problem is considered as NP-complete even if $\omega_{ij} = 1$ for all edges $i, j = 1, \dots, n$ \cite{kap}.
   However, by adding a linear perturbation term, this problem can be solved efficiently  by the canonical duality theory  \cite{wang-ea}.
}
\end{remark}


\subsection{Unified model in mathematical physics}
In analysis and mathematical physics,
the configuration variable $\bchi(t,\bx)$ is a continuous field function
$\bchi : [0,T]\times  \Oo \subset \real\times  \real^d\rightarrow
\omega \subset \real^p$ (which is a hyper-surface if $d+1 = p$ in differential geometry).
The linear operator $\DD = ( \partial_t, \partial_x)$ is a partial differential operator
and the stored energy $\WW(\DD\bchi) = \TT(\partial_t \bchi) - \UU(\partial_x \bchi)$ with $\TT(\bv)$ as the kinetic energy
and $\UU(\beps)$ as deformation energy.
Since $\bv = \partial_t \bchi$ is a vector, the objectivity for  kinetic energy $\TT(\bv)$ is also known as isotropy.
But $\beps = \partial_x \bchi$ is a tensor,  the  deformation energy $\UU(\beps)$  should be an objective function.
In this case, the  Gao and Strang  model (\ref{eq-gs}) is
\eb
\min \left\{ \PP(\bchi) = \TT(\partial_t \bchi) - \UU(\partial_x \bchi) - \la \bchi , \barbchi^* \ra \; | \; \bchi \in \calX_c \right\} .
\ee
The stationary condition $\delta \PP(\bchi) = 0$ leads to a general nonlinear  partial differential equation
\eb
\partial^*_t \partial_{\bv}  \TT(\partial_t \bchi) - \partial^*_x \partial_{\beps} \UU(\partial_x \bchi) = \barbchi^*.
\ee
The nonlinearity of this equation mainly depends on $\TT$ and $\UU$.
For Newtonian mechanics, $\TT(\bv)$ is quadratic. By the objectivity, the deformation energy $\UU(\beps)$ can also
 be split into quadratic part and a nonlinear part  such that
$\WW(\DD \bchi) = \half \la \bchi, \bQ \bchi \ra + \VV(\bD \bchi)$, where $\bQ:\calX_c \rightarrow \calX^* $ is a self-adjoint operator,
$\bD$ is a linear operator, and $\VV(\beps)$ is a nonlinear objective functional.
The most simple example   is a fourth-order polynomial
\eb
\VV(\beps ) = \int_\Oo \half \left( \half \| \beps \|^2 - \lam \right)^2 \dO,
\ee
 which is nonconvex for  $\lam > 0$. This nonconvex functional appears extensively in mathematical physics. 
In fluid mechanics and thermodynamics, $\VV(\beps )$ is the well-known {\em  van de Waals double-well energy}. 
  It is also  known as the {\em sombrero potential} in   cosmic string theory  \cite{edm-cop-kib}, or the {\em Mexican hat} in {\em   Higgs mechanism} \cite{davier}
 and  quantum field theory \cite{a-e}. 
For  this most simple nonconvex potential, the general model (\ref{eq-gs}) can be written as
\eb
\bQ \bchi + \bD^*  \left[  \left( \half \|\bD \bchi \|^2 - \lam \right) \bD \bchi \right]   = \barbchi^* . \label{eq-gsg}
\ee
This model covers many well-known equations.

1) {\bf  Duffing equation} ($\bQ = - \partial^2_t$ and    $\bD = \bI$ is an identical operator):
\eb
\chi_{_{tt}} +    \left( \half \chi^2 - \lam \right) \chi = f (t)
\ee

2) {\bf Landau-Ginzburg equation} ($\bQ = -\Delta, \; \bD = \bI$):
\eb
-\Delta \bchi +    \left( \half \|\bchi\|^2 - \lam \right) \bchi = \bff
\ee

3) {\bf Cahn-Hillar equation} ($\bQ  = -\Delta + \curl \curl, \;\; \bD = \bI$):
\eb
-\Delta \bchi + \curl \curl \bchi +   \left( \half \|\bchi\|^2 - \lam \right) \bchi = \bff.
\ee

4) {\bf Nonlinear Gorden equation} ($\bQ = - \partial_{tt} + \Delta, \;\; \bD = \bI$):
\eb
-\chi_{_{tt}} + \Delta \bchi +  \left( \half \|\bchi\|^2 - \lam \right) \bchi = \bff.
\ee

5) {\bf Nonlinear Gao beam} ($\bQ =  \rho \partial_{tt} + K \partial_{xxxx} , \;\; \bD = \partial_x$):
\eb
\rho \chi_{_{tt}} + K \chi_{_{xxxx}} - \left[\left(\half \chi_{_x}^2 - \lam \right)\chi_{_x} \right]_x= f,
\ee
where $\lam \in \real$ is an axial force and $f(t,x)$ is the lateral load.

According to the nonlinear classification discussed in Section \ref{sec-cdt}, the general equation (\ref{eq-gsg})
is semilinear as long as $\bD = \bI$.
While the nonlinear Gao beam is quasi-linear.
However,  if $\lam > 0$, all these PDEs equations are geometrically nonlinear but physically linear
since by  the canonical transformation
\[
\bxi = \Lam(\beps) = \half \| \beps \|^2 - \lam , \;\; \VV(\beps) = \Phi(\Lam(\beps) ) = \int_\Oo \half \bxi^2 \dO ,
\]
the canonical  duality relation $ \bxi^* = \partial \Phi(\bxi) = \bxi $ is linear.
  {\small  \begin{figure}[h]
\includegraphics[scale=.45]{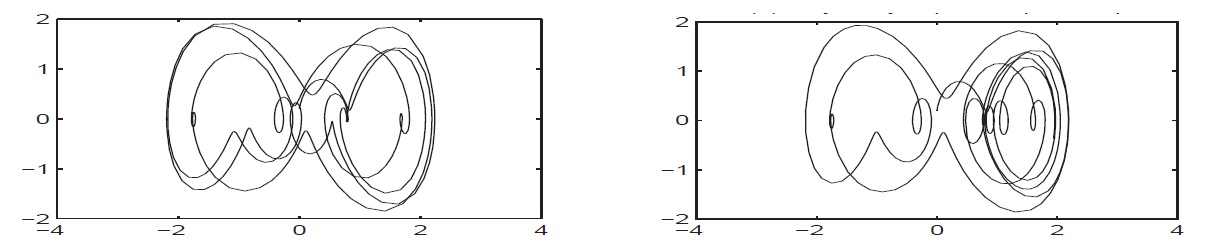} 
 \caption{ Chaotical trajectories of the nonlinear Gao beam computed by ``ode23'' (left) and ``ode15s'' (right) in MATLAB} \vspace{-.2cm}\label{fig-chaos}
 \end{figure}
 }

The geometrical nonlinearity represents large deformation in continuum physics, or far from the equilibrium state in complex systems, which is necessary for nonconvexity but not sufficient.
The nonconvexity of a geometrically nonlinear problem depends on  external force and internal  parameters.
For example,  the total potential of  the nonlinear Gao  beam is nonconvex only if the
compressive load $\lam > \lam_c$, the Euler buckling load, i.e. the first eigenvalue of $K\chi_{_{xxxx}}$ \cite{gao-mrc96,gao-beam00}. In this case, the two minimizers represent the two buckled states,
while the local maximizer represents the unbuckled (unstable) state.
For dynamical loading,
 these two local minimizers are very sensitive to the driving force and initial conditions
  this nonconvex  beam model could produce chaotic vibration.
 The so-called strange attractor is actually a  local minimizer
\cite{gao-thes02,gao-amma03}.
 Particularly, if the variable $\chi(t,x)$ can be separate variable as $\chi=q(t) \sin(\theta x)$, this nonlinear beam model is equivalent to
the Duffing equation, which is well-known in chaotic dynamics.
Figure \ref{fig-chaos} shows clearly that for the same given initial data,
  the same Runger-Kutta iteration but with different solvers in MATLAB produces very different
 ``trajectories'' in phase space $q$-$p$ ($p=q_{,t}$).
 Therefore, this nonlinear beam model is important for understanding  many challenging   problems
 in both mathematics and engineering applications and has been subjected to extensive study recently  \cite{ahn,bengue-etal,cai-gao-qin,gao-thes02,kuttler-etal,li-etal,macha-netu}.

The canonical duality theory has been successfully for modeling real-world problems in
nonconvex/nonsmooth dynamical systems \cite{gao-phil01}, differential geometry \cite{gao-yang-sam},
contact mechanics \cite{gao-jmaa98},
post-buckling structures \cite{gao-beam00},
multi-scale phase transitions of solids  \cite{gao-yu},  and general mathematical physics (see Chapter 4, \cite{gao-dual00}).

\section{Applications in large  deformation mechanics}
 For mixed boundary-value problems,
 the input $\barbchi^* $ is  the body force $ \bff  $ in the domain $\Oo\subset \real^d $
and surface traction $\bt$ on the boundary $\Gt\subset \partial \Oo$.
The external energy
\eb
\FF(\bchi) = \la \bchi, \barbchi^* \ra = \int_\Oo \bchi \cdot \bff  \dO
+ \int_{\Gt} \bchi \cdot \bt \dG
\ee
is a linear functional defined on
$\calX_a = \{ \bchi \in \calC^1[\Oo; \real^p] | \;\; \bchi  = 0 \; \mbox{ on } \Gu\}$.
For a hyper-elastic material deformation problem, we have $\dim \Oo = d = p =3$. The stored energy  $W(\bF)$
 is usually a nonconvex functional of the deformation gradient tensor
  $\bF = \nabla \bchi $
  \eb
  \WW(\bF) = \int_\Oo \UU(\bF ) \dO,
  \ee
  where $\UU(\bF )$ is the stored energy density defined on $\calW_a =
{\mathbb{M}}_+^3 = \{ \bF= \{ F^i_\alp\} \in \real^{3\times 3} | \;\; \det \bF > 0 \}$.
Thus, on the kinetically admissible space
\[
\calX_c = \{ \bchi \in  \calC^1[\Oo; \real^d] | \;\;\det ( \nabla \bchi ) > 0 ,\;\; \bchi  = 0 \; \mbox{ on } \Gu\},
\]
the general model  (\ref{eq-gs}) is a typical nonconvex variational problem
\eb
\min_{\bchi \in \calX_c }  \left\{ \PP(\bchi) = \int_\Oo \UU(\nabla \bchi) \dO - \int_\Oo \bchi \cdot \bff  \dO
- \int_{\Gt} \bchi \cdot \bt \dG  \right\} . \label{eq-nela}
\ee
The linear operator
$D = \grad : \calX_a \rightarrow {\mathbb{M}}_+^3  $ in this problem  is a gradient.
The stationary condition $\delta \Pi(\bchi) = 0$  leads to  a mixed boundary-value  problem (BVP)
\eb
(BVP): \;\; \; A(\bchi) = \nabla^* \partial_\bF \WW(\nabla \bchi) = \left\{
\begin{array}{ll}
- \nabla \cdot  \nabla_\bF \UU(\nabla \bchi)  = \bff \;\; \mbox{ in } \Oo, \label{eq-pde}\\
\bn \cdot \nabla_\bF \UU(\nabla \bchi) = \bt \;\; \mbox{ on } \Gt .
\end{array}
\right.
\ee
According to the definition of  nonlinear PDEs, the first equilibrium equation (\ref{eq-pde}) is fully nonlinear
as long as $\partial \UU(\bF)$ is nonlinear.
However, it is  geometrically linear if $\UU(\bF)$ is convex. It is completely nonlinear only if $\UU(\bF)$ is nonconvex.
Therefore, the definition of fully nonlinearity in PDEs can't be used to identify difficulty of the nonlinear problems.

It is well-known in finite deformation theory that the convexity of the stored energy density $\UU(\bF)$
contradicts the most immediate physical experience (see Theorem 4.8-1,  \cite{ciarlet1}).
Indeed,  even its  domain ${\mathbb{M}}_+^3$   is not a convex subset of $\real^{3\times 3}$ (Theorem 4.7-4, \cite{ciarlet1}).
Therefore,  the solution to the (BVP) is only a stationary point of the total potential $\Pi(\bchi)$.
In order to identify   minimizer of the problem,
many generalized convexities have been suggested
 and the following results are well-known (see \cite{gao-dual00}):
\eb
 \UU(\bF) \mbox{ is  convex } \Rightarrow   \mbox{ poly-convex } \; \Rightarrow
 \; \mbox{ quasi-convex}\footnote{The quasiconvexity used in variational calculus and continuum physics  has an  entirely different meaning from that used in optimization, where a function $f:\real^n \rightarrow \real$ is called  quasiconvex if 
its level set ${\cal L}_{\alp}[f] = \{ x \in \real^n | \; f(x) \le \alp \} $ is convex. For example, the nonconvex function $f(x) = \sqrt{|x|}$ is quasiconvex.}
 \; \Rightarrow   \mbox{ rank-one convex}.
\ee
 If $\UU \in {\cal C}^2(\mathbb{M}_+^3)$, then the
rank-one convexity is equivalent to the Legendre-Hadamard (L.H.) condition:
\eb
\sum_{i,j = 1}^3 \sum_{\alp,\beta =1}^3 \frac{\partial^2 \UU(\bF) }{\partial F^i_\alp \partial F^j_\beta} a_i a_j b^\alp b^\beta \ge 0 \;\; \forall {\bf a} = \{ a_i \}  \in \real^3 , \; \forall {\bf b} = \{ b^\alp \} \in \real^3.
\ee
The   Legendre-Hadamard condition in finite elasticity
is also referred to as the {\em  ellipticity condition}, i.e.,
if the L.H. condition holds, the partial differential operator $A(\bchi)$ in (\ref{eq-pde})
 is considered to be elliptic. %
For one-dimensional problems $\Oo \subset \real$, all these convexities are equivalent  and the rank-one convexity is the
well-known convexity in vector space.
We should emphasize that  these generalized convexities and L.H. condition are local criteria not global.
As long as the total potential  $\Pi(\bchi)$ is locally nonconvex in certain domain of $\Oo$,
 the  boundary-value problem (\ref{eq-pde}) could have  multiple solutions $\bchi(\bx)$
at each material point $\bx \in \Oo$ and the total potential $\Pi(\bchi)$ could have infinitely number of local minimizers
(see \cite{gao-ogden-qjmam}). This is the main difference between nonconvex analysis and nonlinear PDEs,
which is a key point to understand NP-hard problems in computer science and global optimization.
Unfortunately, this difference is not fully understood in both fields.
It turns out that extensive efforts have been devoted
 for solving   nonconvex variational problems directly.
 It was discovered by Gao and Ogden in 2008 that  even for one-dimensional problems,
 the L.H. condition can only identify local local minimizers, and
 a geometrically nonlinear ODE  could have infinite number solutions,
  both local and global minimal solutions
 could be nonsmooth and can't be determined by any Newton type of numerical methods  \cite{gao-ogden-qjmam}.


By the objectivity of  the stored energy density $\UU(\bF)$, it is reasonable to assume a canonical function $\VV(\bC)$ such that
the following canonical transformation holds:
\eb
\WW(\bF) = \Phi(\Lam(\bF)) = \int_\Oo \VV(\bF^T \bF) \dO .
\ee
In this transformation, the geometrical nonlinear operator $\Lam(\bF) =\bF^T \bF$ is quadratic (objective) and
  $\bC = \bF^T \bF \in {\mathbb{S}}^+ = \{ \bC = \{ C_{\alp \beta} \} \in \real^{3 \times 3} | \;\; \bC = \bC^T , \;\; \bC \succ 0 \}$
is the well-known right Cauchy-Green strain tensor.
Its canonical dual $\bS = \partial \Phi(\bC) = \nabla \VV(\bC) \in { \mathbb{S}}$ is a {\em second Piola-Kirchhoff type stress tensor}\footnote{The second Piola-Kirchhoff stress tensor
is defined by $\bT = \partial \Phi(\bE)$, where $\bE = \half (\bC - \bI)$ is the Green-St. Venant strain tensor. Therefore, we have $\bS = 2 \bT$.}.
In terms of the canonical strain measure $\bC (\bF)$, the  kinetically admissible space $\calX_c  = \{ \bchi \in \calC^1[\Oo, \real^3] | \;\; \bC(\nabla \bchi)  \in  {\mathbb{S}}^+ , \;\; \bchi = 0 \mbox{ on } \Gu\}$
is convex and the
 nonconvex variational problem (\ref{eq-nela}) can be written in the canonical form
\eb
\min \left \{ \Pi(\bchi) = \Phi(\bC(\nabla \bchi)) -  \la \bchi, \barbchi^* \ra  | \;\; \bchi \in \calX_c \right\}.
\ee
By the Legendre transformation $\VV^*(\bS) = \{ \bC: \bS - \VV(\bC) | \; \bS = \nabla \VV(\bC) \}$,
  the  total complementary functional $\Xi(\bchi, \bS)$ has the following form: 
\eb
\Xi(\bchi, \bS) = \int_\Oo \left [ \bC(\nabla \bchi) : \bS - \VV^*(\bS)  - \bchi \cdot \bff \right]\dO
- \int_\Gt \bchi \cdot \bt \dG. \label{eq-xie}
\ee
By the fact that the linear operator $\DD = \grad$ is a differential operator, it is difficult to find its inverse operator.
In order to obtain  the canonical dual $\Pi^d(\bT)$,
  we need to introduce the following  {\em statically admissible space}
\[
\calT_c = \{ \btau \in \calC^1[\Oo; \real^{3\times 3} ] | \; - \nabla \cdot \btau = \bff \mbox{ in } \Oo, \;\;
\bn \cdot \btau = \bt \mbox{ on } \Gt \}.
\]
Clearly, for any given $\bchi \in \calX_a = \{ \bchi \in \calC^1[\Oo; \real^3] |\; \det (\nabla \bchi) > 0, \;\;
\bchi = 0 \mbox{ on } \Gu \}$,  the external energy $\FF(\bchi)$ can be written equivalently as
\eb
\FF_\btau(\bchi) =  \int_\Oo  \bchi \cdot (-\nabla \cdot \btau )\dO + \int_{\Gt} \bchi \cdot \bt \dG
= \int_\Oo (\nabla \bchi) : \btau \dO\;\; \;\forall \btau \in \calT_c
\ee
Thus,  for any given $\btau \in \calT_c$, the $\Lam$-conjugate of $\FF(\bchi)$
can be obtained
\eb
\FF^\Lam_\btau(\bS) = \sta \{\la \bC(\nabla \bchi) ; \bS \ra - \FF_\btau(\bchi) | \; \bchi \in \calX_a\}
= -\int_\Oo  \frac{1}{4}  \tr  (\btau \cdot \bS^{-1} \cdot \btau^T ) \dO.
\ee
Its domain should be
\eb
\calS_c = \{ \bS \in \calE^*_a | \; \det(\btau \cdot \bS^{-1}) > 0  \}.
\ee
Therefore,   the pure complementary energy can be obtained as
\eb
\Pi^d(\bS; \btau) = -  \int_\Oo\left[ \frac{1}{4}  \tr  (\btau \cdot \bS^{-1} \cdot \btau^T )  +  \VV^*(\bS) \right] \dO  , 
\ee
which depends on not only the canonical stress $\bS \in \calS_c$, but also the statically admissible field $\btau\in \calT_c$.
Let
\eb
\calS_c^+ = \{ \bS \in \calS_c | \; \bS \succ  0   \}, \;\;\; \calS_c^- = \{ \bS \in \calS_c | \; \bS \prec 0   \}.
\ee
\begin{thm}[Pure Complementary Energy Principle, Gao \cite{gao-ima98,gao-mrc99,gao-dual00}]\label{thm-gao}$\;$ \hfill

If $(\barbS, \barbtau)\in \calS_c \times \calT_c$ is a stationary points of $\Pi^d(\bS; \btau)$, then
 the deformation defined by
 \eb
 \barbchi  (\bx) = \half   \int_{\bx_0}^\bx  \barbtau \cdot \barbS^{-1} \mbox{d} \bx \label{eq-solu}
 \ee
along any path from $\bx_0 \in \Gamma_{\chi}$ to $\bx \in \Oo$ is a critical point of $\Pi(\bchi)$ and
$\Pi(\barbchi) = \Pi^d(\barbS; \barbtau)$.
Moreover, $ \barbchi (\bx)$ is a global minimizer of $\Pi(\bchi)$ if $\barbS (\bx)  \in \calS_c^+ \;\; \forall \bx \in \Oo$.

The vector-valued function
$ \barbchi (\bx)$ is a solution to the boundary-value problem 
of the second equilibrium equation in (\ref{eq-pde})  if the compatibility condition
$ \nabla \times (\barbtau \cdot \barbS^{-1} ) = 0  $ holds.
\end{thm}
{\bf Proof}.
Using Lagrange multiplier $\bchi \in \calX_a$ to relax the equilibrium conditions in $\calT_c$, we have
\eb
\Theta (\bS; \btau, \bchi) =  -  \int_\Oo\left[ \frac{1}{4}  \tr  (\btau \cdot \bS^{-1} \cdot \btau^T )  +  \VV^*(\bS) \right] \dO
- \int_\Oo \bchi \cdot(   \nabla \cdot \btau + \bff ) \dO + \int_\Gt \bchi \cdot \bt \dG.
\ee
Its stationary condition leads to
\eb
 2 \nabla \bchi = \btau \cdot \bS^{-1} \label{eq-fts}
\ee
\eb
4 \bS \cdot (\nabla \VV^*(\bS) ) \cdot \bS = \btau^T \cdot \btau \label{eq-cda}
\ee
and the equilibrium equations in $\calT_c$.
From (\ref{eq-fts}) we have $\btau = 2 (\nabla \bchi) \cdot \bS$. Substituting this into (\ref{eq-cda}) we have
$(\nabla \bchi)^T (\nabla \bchi) = \nabla \VV^*(\bS)$, which is equivalent to
$\bS = \nabla \VV(\bC(\nabla \bchi))$ due to the canonical duality.
Thus, from the canonical transformation, we have
\eb
\btau = 2 (\nabla \bchi) \cdot (\nabla_\bC  \VV(\bC(\nabla \bchi))) = \nabla_\bF  \UU(\nabla \bchi) \label{eq-tau}
\ee
due to the chain role. This shows that the integral (\ref{eq-solu}) is indeed a stationary point of $\PP(\bchi) $ since
 $\btau \in \calT_c$.

By the fact that $\bC = \Lam(\bF)$ is a quadratic operator, the Gao-Strang gap function is
\[
G_{ap} (\bchi, \bS) = \int_\Oo \tr[(\nabla \bchi) \cdot \bS \cdot (\nabla \bchi) ] \dO.
\]
Clearly, $G_{ap}(\bchi, \bS)$ is non negative for any given $\bchi \in \calX_a$ if and only if $\bS(\bx) \in \calS^+_c \;\; \forall \bx \in \Oo$.
Replace $\nabla \bchi = \half \btau \cdot \bS^{-1}$, this gap function reads
\[
G_{ap} (\bchi(\bS, \btau), \bS) = \int_\Oo  \frac{1}{4} \tr[\btau\cdot  \bS^{-1}  \cdot \btau^T] \dO,
\]
which is convex for any $\btau\in \calT_c$ if and only if $\bS(\bx) \in \calS_c^+ \;\;\forall \bx \in \Oo$.
Therefore, the canonical dual $\Pi^d(\bS; \btau)$ is concave on $\calS^+_a \times \calT_c$.
By the canonical min-max duality, $\barbchi$ is a unique global minimizer if $\barbS((\bx) \succ 0 \;\; \forall \bx \in \Oo$.

 The compatibility condition $\nabla \times (\btau \cdot \bS) = 0$ is necessary for an analytical solution to the
 mixed boundary-value problem (\ref{eq-pde}) due to the fact that $\curl \;\grad \bchi = 0$. \hfill  $\square $\\

The pure complementary energy principle was first proposed  by Gao (1997)  in   post-buckling problems of a large deformed  beam  \cite{gao-amr97}.
 Generalization to 3-D finite  deformation theory and nonconvex analysis were given during 1998-2000 \cite{gao-ima98,gao-mrc99,gao-mecc99,gao-dual00,gao-na00}.
The equation (\ref{eq-cda}) is  called the {\em canonical dual algebraic equation} first obtained  in 1998 \cite{gao-ima98}.
This equation shows that by the canonical dual transformation, the nonlinear partial differential equation can be equivalently reformed as an algebraic equation.
The equation (\ref{eq-tau}) show that the statically admissible field $\btau = \nabla \UU(\bF)$ is actually the
{\em  first Piola-Kirchhoff stress}.
For one-dimensional problems, $\btau \in \calT_c$ can be easily obtained by the given input.
For   geometrically nonlinear problems, $\nabla \VV^*(\bS)$ is linear and  (\ref{eq-cda})
  can be solved analytically  to obtain a complete set of analytical solutions
\cite{gao-dual00,gao-na00,gao-anti,gao-ogden-qjmam,gao-ogden-zamp}.
  By the triality theory, the positive solution $\bS  \in \calS^+_c$  produces a global minimal solution $\barbchi$, while the
negative  $\bS \in \calS^-_c$ can be used to identify   local extremal solutions.
To see this,   let us consider the Hessian of the stored energy $\UU(\bF)  = \VV(\bC(\bF))$.
By  chain rule,  we have
 \eb
 \frac{\partial^2 \UU(\bF)}{\partial F^i_\alp\partial F^j_\beta} = 2 \delta^{ij}  S_{\alp\beta} + 4
 \sum_{\theta, \nu = 1}^3  F^i_\theta  H_{\theta \alp\beta \nu} F^j_\nu,
 \ee
 where ${\bf H} = \{ H_{\theta \alp\beta \nu}\} = \nabla^2 \VV(\bC) \succ 0 $
due to the convexity of  the canonical function $\VV(\bC)$.
Clearly, if   $\bS \succeq 0 $,  the  L.H. condition holds and the associated
  $\barbchi$ is a global minimal solution.
   By the fact that $2 \bF = \btau \bS^{-1}$, we know that
$\nabla^2 \UU(\bF)$  could be either positive or negative definite even if $\bS \prec 0$.
Therefore, depending the  eigenvalues of $\bS\prec 0$, the L.H. condition could  also hold
at a local minimizer of $\Pi(\bchi)$ \cite{gao-anti}.
 This shows that the triality theory can be used to  identify both global and local extremal solutions,
  while the L.H. condition is only a  necessary condition for a local minimal solution.
  It is known that an elliptic  equation is corresponding to a convex variational problem.
   Therefore, it is a question if the Legendre-Hadamard  condition can still be called as the
  ellipticity condition in finite elasticity and nonconvex analysis.
  By the fact that  the well-known open problem  left by   Reissner {\em et al} \cite{reissner} has been solved by
   Theorem \ref{thm-gao}, the pure complementary energy principle
 is known as  the Gao principle in  literature (see \cite{li-gupta}).

The canonical transformation  $\WW(\bF) = \Phi(\Lam(\bF))$
is not unique since  the geometrical operator $\Lam(\bF)$  can be chosen  differently to have different canonical strain measures.
For example, the well-known {\em Hill-Seth strain family}
\eb
\bE^{(\eta)} = \Lam(\bF) = \frac{1}{2 \eta} [ (\bF^T \cdot \bF)^\eta - \bI ]
\ee
is a geometrically admissible  objective strain measure for any given $\eta \in \real$  (see Definition 6.3.1, \cite{gao-dual00}).
Particularly, $\bE^{(1)} $ is the well-known {\em Green-St. Venant strain tensor}  $\bE$.
For   {\em St. Venant-Kirchhoff materials}, the  stored  strain density  is quadratic: $\VV(\bE) = \half \bE : \bH : \bE $,
  where $\bH$ is the Hooke tensor. Clearly, $\VV(\bE)$ is convex but
\[
\UU(\bF) = \VV(\bE(\bF)) = \frac{1}{8}  (\bF^T \cdot \bF - \bI) : \bH :   (\bF^T \cdot \bF - \bI )
\]
is a (nonconvex) double-well type function of
  $\bF$, which is not even rank-one convex \cite{raou}.
   The canonical duality is linear  $\bT = \nabla \VV(\bE) = \bH : \bE$ and
 the generalized total complementary energy $\Xi(\bchi, \bT)$ is the well-known Hellinger-Reissner complementary energy
\eb
\Xi(\bchi,\bT) =  \int_\Oo \left [ \bE(\nabla \bchi) : \bT - \half \bT : \bH^{-1} : \bT -   \bchi \cdot \bff   \right]\dO - \int_\Gt \bchi \cdot \bt \dG .
\ee
In this case, the primal problem (\ref{eq-nela}) is a  geometrically nonlinear variational problem, and its canonical dual functional is
\eb
 \Pi^d(\bT; \btau) = -  \int_\Oo \half \left[ \tr  (\btau \cdot \bT^{-1} \cdot \btau^T + \bT ) +   \bT : \bH^{-1} : \bT \right]\dO .
\ee
The canonical dual algebraic equation (\ref{eq-cda}) is then a cubic tensor equation
\eb
2 \; \bT \cdot ( \bH^{-1} : \bT + \bI ) \cdot \bT = \btau^T \cdot \btau \label{eq-cdals}
\ee
For a given statically admissible stress field $\btau \in \calT_c$, this tensor equation could have at most 27 solutions $\bT(\bx)$
at each material point $\bx \in \Oo$,
but only one $\bT(\bx) \succ 0$, which leads to a global minimal solution \cite{gao-haji}.

For many real-world problems, the statically admissible stress $\btau \in \calT_c$ can
 be uniquely obtained and the
canonical dual algebraic equation (\ref{eq-cdals}) can be solved to obtain all possible stress solutions.
The  canonical duality-triality theory has been used successfully for solving a class of nonconvex
variational/boundary value  problems  \cite{gao-na00,gao-lv,gao-ogden-qjmam},
  pure azimuthal shear  \cite{gao-ogden-zamp} and
  anti-plane shear problems \cite{gao-anti}.

\section{Applications to computational mechanics and global optimization}
Numericalization for solving  the nonconvex variational problem (\ref{eq-gs}) leads to a global optimization problem
in a finite dimensional space $\calX= \calX^*$.
   In complex systems, the decision variable
 $\bchi$ could be either vector or matrix. In
 operations research, such as logistic and supply chain management sciences,  $\bchi$ can be even a
  high-order matrix $\bchi= \{ \chi_{ij... k}\}$.
  Correspondingly, the linear operator $\DD:\calX_a \rightarrow \calW_a$ is
a matrix or high-order  tensor.
In general global optimization problems, the internal energy $\WW(\DD\bchi)$ is not necessary to be an objective function.
As long as the canonical transformation $\WW(\DD(\bchi) = \Phi(\Lam(\DD\bchi))$ holds, the
 canonical duality-triality theory  can be used   for solving a large class of nonconvex/discrete optimization problems.

 \subsection{Canonical dual finite element method}
It was shown in \cite{gao-jem} that by using independent finite element interpolations for  displacement and
generalized stress:
\eb
\bchi(\bx) = \bN_u (\bx)  \bq^e , \;\; \bS(\bx) = \bN_\vsig (\bx) \bp^e \;\; \forall \bx \in \Oo^e \subset \Oo , \label{mixedFEM}
\ee
the total complementary functional $\Xi(\bchi, \bS)$ defined by  (\ref{eq-xie})  can be discretized as a function in finite dimensional space
\eb
\Xi(\bq, \bp) = \frac{1}{2}  \bq^T \bG(\bp) \bq - \Phi^*(\bp) - \bq^T \bff,
\ee
where $\bff$ is the generalized force and  $\bG(\bp)$ is the Hessian matrix of the discretized Gao-Strang  complementary gap function.
In this case, the pure complementary energy can be formulated explicitly as  \cite{gao-jem}
\eb
\Pi^d(\bp) = - \frac{1}{2} \bff^T \bG^+(\bp) \bff - \Phi^*(\bp),
\ee
where $\bG^+$ represents a generalized inverse of $\bG$. Let
\[
\calS_c^+ = \{ \bp \in \real^m | \; \bG(\bp) \succeq 0 \}, \;\; \calS_c^- = \{ \bp \in \real^m | \; \bG(\bp)  \prec  0 \}.\vspace{-.2cm}
\]
By the fact that $\Pi^d(\bp)$ is concave on the convex set $\calS^+$, the canonical dual FE programming problem
\eb
\max \{ \Pi^d(\bp) | \; \bp \in \calS_c^+ \}
\ee
can be solved easily (if $ \calS_c^+ \neq \emptyset$) to obtain the global maximizer $\barbp$. By the triality theory, we know that $\barbq = \bG^+(\barbp) \bff$ is a global minimizer of
the nonconvex potential $\Pi(\bq)$.
On the other hand, if $\dim \bq = \dim \bp$, the biggest local min and local max of $\Pi(\bq)$ can be obtained respectively
 by \cite{gao-wu-jimo} 
\[
\min \{ \Pi^d (\bp) | \; \bp \in \calS_c^- \}, \;\; \max \{ \Pi^d(\bp) | \; \bp \in \calS_c^- \}.
\]

\begin{figure}[t]
 \includegraphics[width=2.7in,height=4cm]{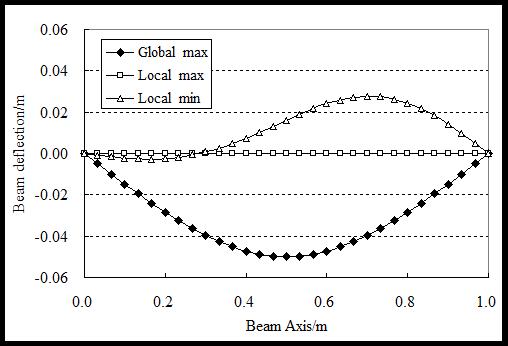}\hspace{1.cm}
 \includegraphics[width=2.7in,height=4cm]{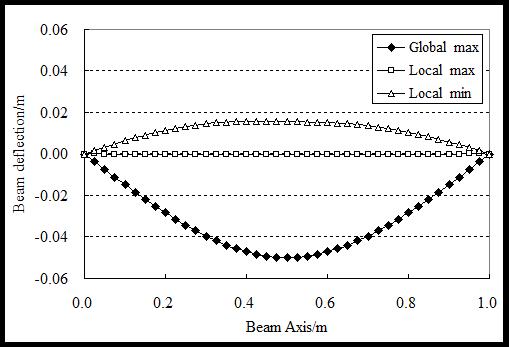}
$\;$\\
\hspace{1cm} $\;\;\;\;\;\;\;\;\; \mbox{  (a) 30 elements.}$  \hspace{5cm} (b) 40 elements.
 \caption{{\small Canonical dual FEM solutions for post-buckled nonlinear beam:
 Global minimal solution, i.e. stable buckled state (doted); local min, i.e. unstable buckled state (triangle);
  and local max, i.e. unbuckled state (squared).}} \label{fig-beam}
   \end{figure}

The canonical dual FEM has been used successfully in phase transitions of solids \cite{gao-yu} and in post-buckling analysis
 for the large deformation beam model (2)
to obtain all three possible solutions  \cite{santos-gao} (see Fig. \ref{fig-beam}).
It was discovered  that the local minimum  is very sensitive to the lateral load and the size of the finite element meshes
(see Fig.  \ref{fig-beam}). This method can be used for solving general nonconvex mechanics problems.

\subsection{Global optimal solutions for  discrete nonlinear dynamical systems}
General nonlinear dynamical systems can be modeled as a nonlinear initial-value problem
\eb
\bchi'(t) = \bF(t, \bchi(t) ) \;\; t \in [0, T] , \;\; \bchi(0) = \bchi_0,\label{eq-nivp}
\ee
where $T > 0$,  $\bF:[0,T]\times \real^d \rightarrow \real^d $ is a given vector-valued function.
Generally speaking, if a nonlinear equation has multiple solutions at each time $t$ in a subset of its domain $[0,T]$,
then the associated initial-valued problem should have infinite number of solutions since the unknown  $\bchi(t) $
is a continuous function.
With time step size $h = T/n$, a discretization of the configuration $\bchi(t)$ is
$\bX = ( \bchi_0, \bchi_1, \cdots, \bchi_n) \in \calX_a \subset  \real^{d \times (n+1)}$.
By  the finite difference method, the initial value problem (\ref{eq-nivp})
can be written approximately as
\eb
\bchi_{k} = \bchi_{k-1} + \half h    \bF(t_{k-1}, \bchi_{k-1}), \;\; k = 1, \cdots, n. \label{eq-fdm}
\ee
This is still a nonlinear algebraic system.
 Clearly, any linear iteration can only produce one of the infinite number solutions, and such a
   numerical ``solution" is very sensitive to
 the step-size and numerical errors. This is the reason why different numerical solvers produce  totally different results,
 i.e. the so-called chaotic solutions.
Rather than the traditional  linear iteration  from an initial value, we use the least squares method such that the
nonlinear algebraic system (\ref{eq-fdm}) can be equivalently written as
\eb
\min_{  \bX \in \calX_a } \left\{ \Pi(\bX) = \half \sum_{k=1}^n \left\| \bchi_k - \bchi_{k-1} -   h   \bF(t_{k-1}, \bchi_{k-1}) \right\|^2 \;   \right\}.\label{eq-lsq}
\ee
Clearly, for any given nonlinear function $\bF(t, \bchi(t))$, this is a  global optimization problem, which could have multiple
minimizers at each $\bchi_k$.  Particularly,
 if $\bF(t,\bchi)$ is quadratic,
then $\Pi(\bX)$ is a double-well typed  fourth order polynomial function, and is considered to be NP-hard in global optimization
even for  $d = 1$ (one-dimensional systems) \cite{a-g-y,sax}.
However, by simply using the quadratic geometrical operator
$\bxi_k = \Lam(\bchi_k) = \bF(t_k, \bchi_k)$, the nonconvex leas squares  problem (\ref{eq-lsq})  can be solved by the canonical duality-triality
theory to obtain global optimal solution.
Applications have been given to the logistic map   \cite{li-zhou-gao} and population growth problems \cite{ruan-gao-ima}.


\subsection{Unconstrained nonconvex minimization}
The general model (\ref{eq-gs})
for unconstrained global optimization can be
written in the following form
\eb
\min \left\{ \PP(\bchi) = \WW(\bD \bchi) + \half \la \bchi , \bA \bchi \ra - \la \bchi, \bff \ra | \;\; \bchi \in \calX_a \right\}, \label{eq-gop}
\ee
where $\bD : \calX_a \rightarrow \calW_a$ and $\bA = \bA^T $ are two  given  operators  and
$\bff \in \calX_a$ is a given input.
For the nonconvex function $\WW(\bw)$, we assume that the canonical transformation $\WW(\bD \bchi) = \Phi(\Lam(\bchi))$
holds for a quadratic operator
\eb
\Lam (\bchi ) =  \left\{ \half  \bchi^T  \bH_{\alp\beta} \bchi  \right\}  :\calX_a \rightarrow \calE_a \subset \real^{m\times m},
\ee
where $\bH_{\alp\beta} =  \bH^T_{\alp\beta} \;\; \forall \alp, \beta \in I_m=\{1, \dots, m\}$ is a linear   operator such that $\calE_a$ is either a vector ($\beta = 1$) or tensor ($\alp, \beta > 1$) space.
By the convexity of the canonical function $\Phi:\calE_a \rightarrow \real$, the
canonical duality $\bS = \partial \Phi(\bxi) \in \calE^*_a \subset \real^{m\times m}$ is invertible and the total complementary function $\Xi :\calX_a \times \calE_a^* \rightarrow \real$
reads
\eb
\Xi(\bchi, \bS) =   \half \la \bchi , \bG(\bS) \bchi \ra - \Phi^*(\bS)    - \la \bchi, \bff \ra
\ee
where
$ \bG(\bS) = \bA +  \sum_{\alp,\beta \in I_m}  \bH_{\alp\beta}  S_{\alp\beta} $.
Thus, on $\calS_c^+ = \{ \bS \in \calE^*_a | \;\; \bG(\bS) \succ 0 \}$,
 the canonical dual problem (\ref{eq-cdmax}) for the unconstrained global optimization reads
\eb
\max \left\{ \Pi^d(\bS) = - \half \la \bG^{-1}(\bS) \bff , \bff \ra - \Phi^*(\bS) \; | \; \; \bS \in \calS_c^+ \right\} .
\ee
The canonical duality-triality theory has been used successfully for solving the following nonconvex problems.

1) {\bf Euclidian Distance Geometry Problem}
\eb
\WW(\bD \bchi) = \sum_{i, j= 1}^n \omega_{ij} \left[ \| \bchi_i - \bchi_j \|^2  - d_{ij} \right]^2 ,
\ee
where the decision variable $\bchi_i \in \real^d$ is a position (location) vector,
$\omega_{ij}, \; d_{ij} >  0 \; \forall  i,j  = 1, \dots, n, \; i \neq j $ are given weight and distance  parameters, respectively.
 The linear operator $\bD \bchi = \{ \bchi_i - \bchi_j \}$ in this problem is similar to the finite difference in numerical analysis.
 Such a problem appears frequently in   computational biology \cite{z-g-y},  chaotic dynamics \cite{li-zhou-gao,ruan-gao-ima},
 numerical algebra \cite{r-g-j},
sensor localization \cite{LV14,ruan-gao-pe}, network communication \cite{gao-ruan-pardalos},  transportation  optimization,
as well as finite element analysis of structural mechanics \cite{cai-gao-qin,gao-yu}, etc.
 These problems are considered to  be NP-hard even the Euclidian dimension $d = 1$  \cite{a-g-y}.
 However, by the combination of the  canonical duality-triality theory and perturbation methods,
  these problems can be solved efficiently (see \cite{ruan-gao-pe}).

  2) {\bf Sum of Fractional Functions}
  \eb
  \WW(\bD \bchi) = \sum_{i\in I_m} \frac{G_i(\bD_g \bchi)}{H_i(\bD_h\bchi)}
  \ee
  where $G_i$ and $H_i > 0  \;\; \forall \; i \in I_m $ are given functions, $\bD_g$ and $\bD_h$ are linear operators.

  3) {\bf Exponential-Sum-Polynomials}
\eb
W(\bD \bchi) =\sum_{i\in I_m} \exp\left(\frac{1}{2}\bchi^T \bB_i\bchi-\alp_i\right) +
\sum_{j\in I_p}\half \left(\frac{1}{2}\bchi^T \bC_j\bchi -\beta_j \right)^2,
\ee
where $\bB_i $ and $\bC_j$ are given symmetrical matrices in $\real^{n\times n}$, $\alp_i, \beta_j $ are given parameters.

4) {\bf Log-Sum-Exp Functions}
\eb
\WW(\bD\bchi) = \frac{1}{\beta} \log \left[ 1 + \sum_{i\in I_p} \exp \left( \beta \left ( \half \bchi^T \bB_i \bchi + d \right) \right) \right],
\ee
where $\beta > 0$,  $\bB_i = \bB^T_i$, and $d \in \real$ are given.

All these  functions appear extensively in modeling  real-world problems, such as computational biology \cite{z-g-y},
 bio-mechanics,  phase transitions \cite{gao-ogden-qjmam},  filter design \cite{wu-gao-teo},   location/transportation and networks optimization  \cite{gao-ruan-pardalos,ruan-gao-pe},  communication
and information theory (see \cite{la-gao}) etc.
By using the canonical duality-triality theory, these problems can be solved nicely (see \cite{chen-gao-jogo,gao-ruan-mmor,LS14,mora-gao-memo,z-g-y-amc}).

\subsection{Constrained global optimization}\label{sec-const}
Recall  the standard mathematical model in global optimization
(\ref{eq-go})
\eb\label{eq: original problem}
\min f(\bx) , \;\; \mbox{ s.t. } \; h_i(\bx) = 0, \;\;  g_j(\bx) \le 0  \;\; \forall i \in I_m , \;\; j  \in I_p,
\ee
where $f$, $g_i$ and $h_j$ are differentiable, real-valued functions  on a subset of $\real^n$ for all $i\in I_m$ and
$j\in I_p$. For notational convenience,
we use vector forms for  constraints
$$
\bg(\bx)=\left(g_1(\bx),\dots, g_m(\bx)\right) , \;\;
\bh(\bx)=\left(h_1(\bx),\dots, h_p(\bx)\right).
$$
Therefore, the feasible space can be defined as
$$
{\cal X}_c:=\{\bx\in \real^n| \bg(\bx)\le 0, \;\; \bh(\bx)=0 \}.
$$

 Lagrange  multiplier method  was originally proposed  by J-L Lagrange from analytical mechanics
 in 1811 \cite{lagrange}.
 During the past two hundred  years, this method and the associated Lagrangian duality theory have been well-developed
 with extensively applications  to many fields of physics,  mathematics and engineering sciences.
 Strictly speaking, the Lagrange multiplier method can be used only for equilibrium constraints.
 For inequality constraints, the well-known KKT conditions are involved.
 Here we  show that both the classical Lagrange multiplier method and the KKT theory can be unified by the canonical duality theory.

 For  convex constrained problem, i.e. $f(\bx)$,  $\bg(\bx)$ and $\bh(\bx)$ are convex,
 the standard canonical dual transformation can be used.
We can choose the geometrical operator $\bxi_0 = \bLam_0 (\bx)= \{ \bg(\bx), \bh(\bx) \} : \real^n \rightarrow \real^{m+p}$ and
let
\[
\Phi_0(\bxi_0) = \Psi_g(\bg) + \Psi_h(\bh),
\]
where
\[
\Psi_g(\bg)= \{
0 \;\;  \mbox{if } \bg\le0, \; \;
+\infty \;\; \mbox{otherwise}\}, \;\;\;
\Psi_h(\bh)= \{
0 \;\;  \mbox{if } \bh=0, \;\;\;
+\infty \;\; \mbox{otherwise} \},
\]
are the so-called indicator functions for the inequality and  equality constraints.
Then the  convex  constrained problem (\ref{eq: original problem})
can be written in the following canonical form
\begin{equation}\label{eq: indicator f}
\min  \left\{ \Pi(\bx) = f(\bx)+ \Phi_0(\bLam_0(\bx)) | \;\; \forall \bx \in \mathbb{R}\right\}.
\end{equation}

By the fact that the canonical function $\Phi_0(\bxi_0)$ is convex and lower semi-continuous, the canonical duality relations
(\ref{eq-cdr}) should be replaced by the following subdifferential forms \cite{gao-jogo00}:
\eb
\bxi_0^*  \in \partial \Phi_0(\bxi_0) \;\; \Leftrightarrow \;\; \bxi_0 \in \partial \Phi_0^*(\bxi^*_0)
\;\; \Leftrightarrow \;\; \Phi_0(\bxi_0) + \Phi_0^*(\bxi_0^*) = \bxi_0^T \bxi^*_0 ,
\ee
where $\Phi_0^* (\bxi^*_0) = \Psi^*_g(\blam) + \Psi_h^*(\bmu)$ is the Fenchel   conjugate of $\Phi_0(\bxi_0)$
and  $\bxi_0^* = (\blam , \bmu)$. By the Fenchel transformation, we have
\[
\Psi_g^*(\blam )=\sup_{\bg \in \real^m} \{\bg^T\blam -\Psi_g(\bg) \}=
\left\{
\begin{array}{ll}
0 & \mbox{if }  \blam \ge 0 \\
+ \infty  & \mbox{otherwise} ,
\end{array}
\right.
\]
\[
\Psi_h^*(\bmu)=\sup_{\bh \in \real^p}\{\bh^T\bmu-\Psi_h(\bh) \} = 0 \;\; \; \forall \;  \bmu\in\real^p.
\]
 It is easy to verify that
for the indicator $\Psi_g(\bg)$,  the canonical duality  leads to
\begin{equation}\label{eq: conditions on g}
\begin{array}{cll}
\blam  \in \partial \Psi_g(\bg) &\Longrightarrow& \blam\ge 0  \\
\bg \in\partial \Psi^*_g(\blam) &\Longrightarrow & \bg \le0  \\
\blam^T \bg =\Psi_g(\bg )+\Psi^*_g(\blam)&\Longrightarrow & \blam^T \bg=0 ,
\end{array}
\end{equation}
which are the KKT  conditions for the inequality constrains $\bg(\bx) \le 0$.
While for $\Psi_h(\bh)$, the canonical duality   lead to
\begin{equation}\label{eq: conditions on h}
\begin{array}{cll}
\bmu \in \partial \Psi_h(\bh ) &\Longrightarrow& \bmu\in \real^p\\
\bh \in\partial \Psi^*_h(\bmu) &\Longrightarrow &  \bh =0  \\
\bmu^T \bh =\Psi_h(\bh)+\Psi^*_h(\bmu)&\Longrightarrow & \bmu^T \bh=0 .
\end{array}
\end{equation}
From the second and third conditions in the (\ref{eq: conditions on h}), it is clear that in order to enforce   the  constrain $\bh(\bx)=0$,
 the dual variable  $\bmu = \{\mu_i\} $ must be not zero   $\forall i\in I_p$.
 This is a special  complementarity condition for equality constrains,
  generally not mentioned  in many textbooks.
However, the implicit constraint  $\bmu \neq 0$
is important in nonconvex optimization.

 By using the Fenchel-Young equality $ \Phi_0(\bxi_0) = \bxi_0^T \bxi_0^* -  \Phi_0^*(\bxi_0^*)$
  to replace $\Phi_0(\bLam_0(\bx))$  in  (\ref{eq: indicator f}),
   the
   {  total complementarity function }   can be obtained in the following form:
\begin{equation}\label{eq: complete complementarity}
\Xi_0(\bx,\bxi^*_0)=f(\bx)+[\blam^T \bg(\bx)-\Psi_g^*(\blam)]+[ \bmu^T \bh(\bx)-\Psi_g^*(\bmu)].
\end{equation}

Let $\bsig_0 = (\blam , \bmu) $. The
dual feasible spaces  should be defined as
\[
{\cal S}_0 = \{ \bsig_0 = ( \blam, \bmu)  \in \real^{m \times p }|\;\; \lambda_i\ge 0 \;\; \forall i\in I_m, \;\;
 \mu_j\neq 0  \;\;\forall j \in I_p\}.
\]
Thus, on the feasible space $\real^n\times{\cal S}_0$,
 the total complementary function (\ref{eq: complete complementarity}) can be simplified as
\begin{equation}\label{eq: lagrangian}
\Xi_0(\bx,\bsig_0)=f(\bx)+\blam^T \bg(\bx) +\bmu^T \bh(\bx)={\cal L}(\bx,\blam , \bmu) ,
\end{equation}
which is   the classical Lagrangian and we have
\[
P(\bx) = \sup \left\{ \Xi_0(\bx, \bsig_0) | \; \forall \bsig_0  \in {\cal S}_0 \right\} .
\]
This shows that the canonical duality theory is an extension of the Lagrangian theory
(indeed,  the total complementary function was called the extended Lagrangian in \cite{gao-dual00}).

For nonconvex constrained problems, the so-called
{\em  sequential  canonical transformation  }  (see Chapter 4, \cite{gao-dual00})
\[
\bLam_0(\bLam_1(\dots (\bLam_k(\bx)) \dots ))
\]
can be used for  target function and constraints to obtain high-order canonical dual problem.
Applications have been given to the high-order polynomial optimization \cite{gao-jogo06,gaot-ima},
nonconvex analysis \cite{gao-dual00}, neural network \cite{la-gao}, and nonconvex constrained problems \cite{g-r-s,gao-yang-opt,lato-gao-opl,mora-gao-jogo,z-g-y-opl}.

\subsection{SDP relaxation  and canonical primal-dual algorithms }
Recall the primal problem $(\calP)$ (\ref{eq-canp})
\[
(\calP): \;\; \min \{ \Pi(\bchi) = \Phi(\Lam(\DD\bchi)) - \la \bchi, \barbchi^* \ra | \; \bchi \in \calX_c\}
\]
and its canonical dual $(\calP^d)$  (\ref{eq-cdmax})
\[
(\calP^d): \;\;\; \max \left\{ \Pi^d(\bS) =  -G^*_{ap}(\bS)  - \Phi^*(\bS) \; | \; \bS \in \calS_c^+ \right\},
\]
where  $G^*_{ap}(\bS) = \half \la  \bG^{-1}(\bS) \bF(\bS) , \bF(\bS) \ra$ is the pure gap function.
 By the fact that $(\calP^d)$   is a concave maximization on a convex domain $\calS^+_c$,
 this canonical dual
  can be solved easily if $\Pi^d(\bxi^*)$ has a stationary point in $\calS_c^+$.
  For  many challenging (NP-hard)  problems,
  the stationary points $\Pi^d(\bS)$ are usually located on the boundary of $\calS^+_c = \{ \bS \in \calS_c | \;\; \bG(\bS) \succ 0 \}$.
  In this case, the matrix $\bG(\bS)$ is singular and the canonical dual problem could have multiple solutions.
  Two methods can be suggested for solving this challenging case.\\

  {\bf 1) SDP Relaxation}. By using the Schur complement Lemma, the canonical dual problem $(\calP^d)$ can be relaxed as  \cite{z-g-y-amc}
  \eb
 (\calP^r): \;\;  \min \Phi^*(\bS) \;\mbox{ s.t. } \; \left( \begin{array}{cc}
   \bG (\bS) & \bF (\bS) \\
  \bF^T(\bS) & 2 G_{ap}(\bS) \end{array} \right)  \succeq 0 , \;\;   \forall \bS \in \calS_c .
  \ee
  Since $\Phi^*(\bS)$ is convex and the feasible space is closed, this relaxed canonical dual problem has at least one solution $\barbS$.
  The associated $\barbchi = \bG(\barbS)^{-1} \bF(\barbS)$ is a solution to $(\calP)$ only if $\barbS$ is a stationary point of $\Pi^d(\bS)$.
Particularly, if $\Phi^*(\bS) = \la \bQ ; \bS \ra $ is linear, $\bF = 0$, $\bG(\bS) = \bS$, and
\[
\calS_c = \{ \bS \in  {\mathbb{S}}_n | \;\; \la  \bA_i ; \bS \ra =  b_i \; \forall  i\in I_m \}
\]
 is a linear manifold, where ${\mathbb{S}}_n = \{ \bS \in \real^{n\times n} | \; \bS = \bS^T \}$ is a symmetrical $n\times n$-matrix space,
  $\bQ , \; \bA_i \in {\mathbb{S}}_n \;\; i\in I_m$ are given matrices and $\bb = \{ b_i \} \in \real^m$ is a given vector,
   then by the notation  $ \bQ \bullet \bS = \la \bQ ; \bS \ra =\tr(\bQ \cdot \bC) = \bQ : \bC$, the
 relaxed canonical dual problem  can be written as
 \eb
 \min  \bQ \bullet  \bS \;\; \mbox{ s.t. } \; \bS \succeq 0 , \;\;   \bA_i \bullet  \bS   =  b_i , \;\; \forall i \in I_m ,
 \ee
which  is a typical Semi-Definite Programming (SDP) problem in optimization \cite{todd}.
This shows that the popular SDP problem is indeed a special case of the canonical duality-triality theory for solving the general
global optimization problem (\ref{eq-gs}).
The SDP method and algorithms have been well-studied in global optimization.
But this method provides only a lower bound approach for the
global minimal solution to $(\calP)$ if its canonical dual has no stationary point in $\calS^+_c$.
Also, in many real-world applications, the local solutions are also important. Therefore, a second method is needed.\\

 {\bf 2)  Quadratic perturbation and canonical primal-dual algorithm}.
 By introducing a quadratic perturbation,  the total complementary function
   (\ref{eq-xiq}) can be written as
\begin{eqnarray*}
\Xi_{\delta_k} (\bchi, \bxi^*) & = & \Xi(\bchi, \bS) + \half \delta_k \| \bchi - \bchi_k \|^2  \\
&= & \half \la \bchi , \bGd(\bS ) \bchi \ra - \Phi^*(\bS ) - \la \bchi, \bFd(\bS )\ra + \half \delta_k \la \bchi_k, \bchi_k \ra ,
\end{eqnarray*}
where $\delta_k > 0, \;  \bchi_k \; \; k\in I_p $ are  perturbation parameters, $\bGd(\bS) = \bG(\bS) + \delta_k \bI $, $\bFd(\bS) = \bF(\bS) + \delta_k \bchi_k$.
Thus, the original canonical dual feasible space $\calS_c^+$ can be enlarged to
$\calS_{\delta_k}^+ = \{ \bS \in \calS_c | \; \bGd(\bS) \succ 0 \}$.
Using the perturbed total complementary function $\Xi_{\delta_k}$, the perturbed canonical dual problem can be proposed
\eb
(\calP^d_k): \;\;  \max \left\{ \min \{ \Xi_{\delta_k} (\bchi, \bS) |\;\; \bchi \in \calX_a \} | \; \; \bS \in \calS_{\delta_k}^+ \right\}
\ee
Based on this perturbed canonical dual problem, a canonical primal�dual algorithm has been developed \cite{wu-li-gao,z-g-y-amc}. \\

\noindent{\bf Canonical Primal-Dual Algorithm}. Given initial data $\delta_0 > 0$, $\bchi_0 \in \calX_a$,  and error allowance $omega > 0$.
Let $k=1$.
\begin{verse}
1) Solve the perturbed canonical dual problem $(\calP^d_k)$ to obtain $\bS_k \in \calS_{\delta_k}^+$. \\
2) Computer $\barbchi_k = [ \bGd (\bS_k)]^{-1} \bFd(\bS_k)$ and let
\[
\bchi_k = \bchi_{k-1}  + \beta_k (\barbchi_k - \bchi_{k-1}), \;\; \beta_k \in [0,1].
\]\\
3) If $|\Pi(\bchi_k) - \Pi(\bchi_{k-1}) | \le \omega $, then stop, $\bchi_k$ is the optimal solution to $(\calP)$.
Otherwise, let $k = k + 1$, go back to 1).
\end{verse}
 In this algorithm, $\{ \beta_k\} $ are given  the parameters, which change the search directions. Clearly, if
 $\beta_k = 1$, we have $\bchi_k =  \barbchi_k $.
 This algorithm has been used successfully for solving a class of benchmark problems  and sensor network optimization problems
 \cite{ruan-gao-pe,z-g-y-amc}.

 Let $\calS_{\delta_k}^- = \{ \bS \in \calS_c | \; \bGd(\bS) \prec  0 \}$. The combination of  this algorithm with the double-min and  double-max dualities in the
 triality theory can be used for finding local optimal solutions \cite{cai-gao-qin}.

\section{Challenges and breakthrough}
In the history of sciences, a ground-breaking   theory usually has to pass through serious arguments and challenges.
This is duality nature and certainly true for the canonical duality-triality theory,
which   has   benefited from
 recent challenges by    M. Voisei,  C. \Z  and his former student R. Strugariu
in a set of 11  papers. These papers   fall naturally into three interesting  groups.
\subsection{ Group 1: Bi-level duality}
One paper in this group   by  Voisei and \Z  \cite{vz3} challenges
  Gao and Yang's work  for solving
 the following  minimal distance between two  surfaces \cite{gao-yang-opt}
\eb
\min \left\{   \half \| \bx - \by\|^2 \; | \;\; g(\bx) = 0, \;\; h(\by) = 0, \;\; \bx, \by \in \real^n\right\},\label{eq-gy}
\ee
where  $g(\bx)$ is  convex, while $h(\by)$ is a nonconvex function.
By the canonical transformation $h(\by) = \VV(\Lam(\by)) -  \by^T \bff $, the Gao-Strang complementary function
was written in the form of $\Xi(\bx, \by, \bsig_0, \vsig)$, where  $\bsig_0 = \{ \lam, \mu)$ is the first level canonical dual variable, i.e. the  Lagrange multiplier
for $\{ g(\bx) = 0, \; h(\by) = 0\}$, while  $\vsig $ is the second level  canonical dual variable for the nonconvex
 constraint (see equation (11) in \cite{gao-yang-opt}).
Using one counterexample
\eb
g(\bx) = \half (\|  \bx \|^2 -1) , \;\; h(\by) = \half \left( \half \|\by - \bc\|^2 - 1 \right)^2  -
\bff^T (\by - \bc),  \label{eq-vz}
\ee
 with $n=2$ and  $ \bc =  (
1,0) , \;\; \bff = (
\frac{\sqrt{6}}{96} ,  0 )$,
Voisei and \Z proved that
``the main results in Gao and Yang  \cite{gao-yang-opt} are false"
 and they concluded:  ``The consideration of the function  $\Xi$ is useless, at least for the problem
studied in \cite{gao-yang-opt}".

This paper raises up two  issues on  different levels.

The first issue is  elementary: there is indeed a  mistake in Gao and Yang's work, i.e. instead of  $ (\bx, \by,\bsig_0, \vsig)$ used in \cite{gao-yang-opt}, the
variables in the total complementary function $\Xi$ should be  the vectors
$\bchi  = (\bx , \by)$ and $(\bsig_0, \vsig)$  since  $\Xi(\bchi, \bsig_0, \vsig)$  is convex in $\bx$ and $\by$ but may not in
$\bchi$.
 This  mistake has been easily corrected in \cite{mora-gao-jogo}.
Therefore, the duality on  this level is:  opposite to  Voisei and Z\u{a}linescu's  conclusion,
the consideration of the  Gao-Strang total complementary function  $\Xi$
is indeed quite useful for solving the challenging   problem  (\ref{eq-gy}) \cite{mora-gao-jogo}.

\begin{figure}[h]
  \centering
 {\label{2DPert64}\includegraphics[width=0.3\textwidth]{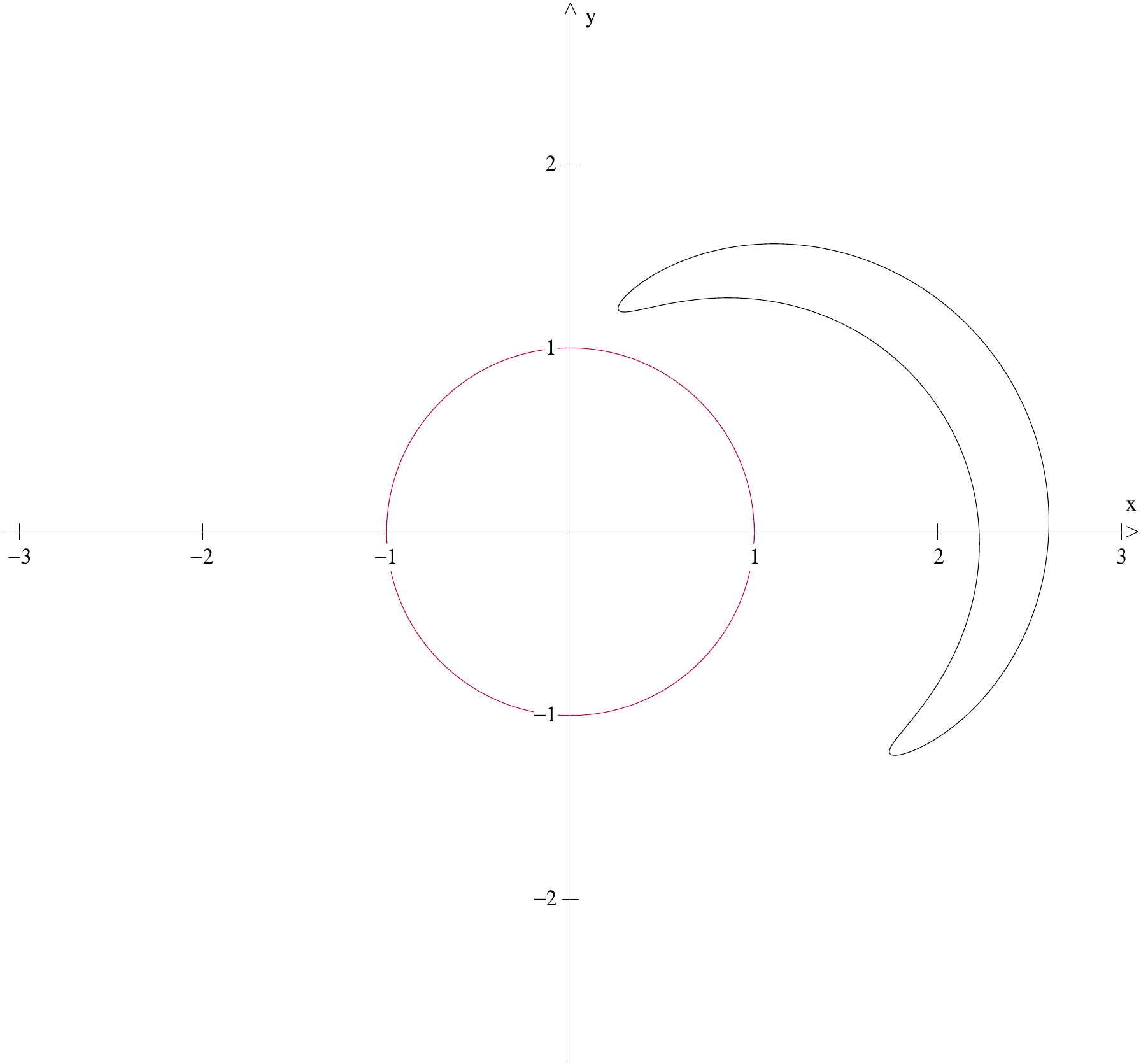}}\quad
 {\label{2DPert100K}\includegraphics[width=0.3\textwidth]{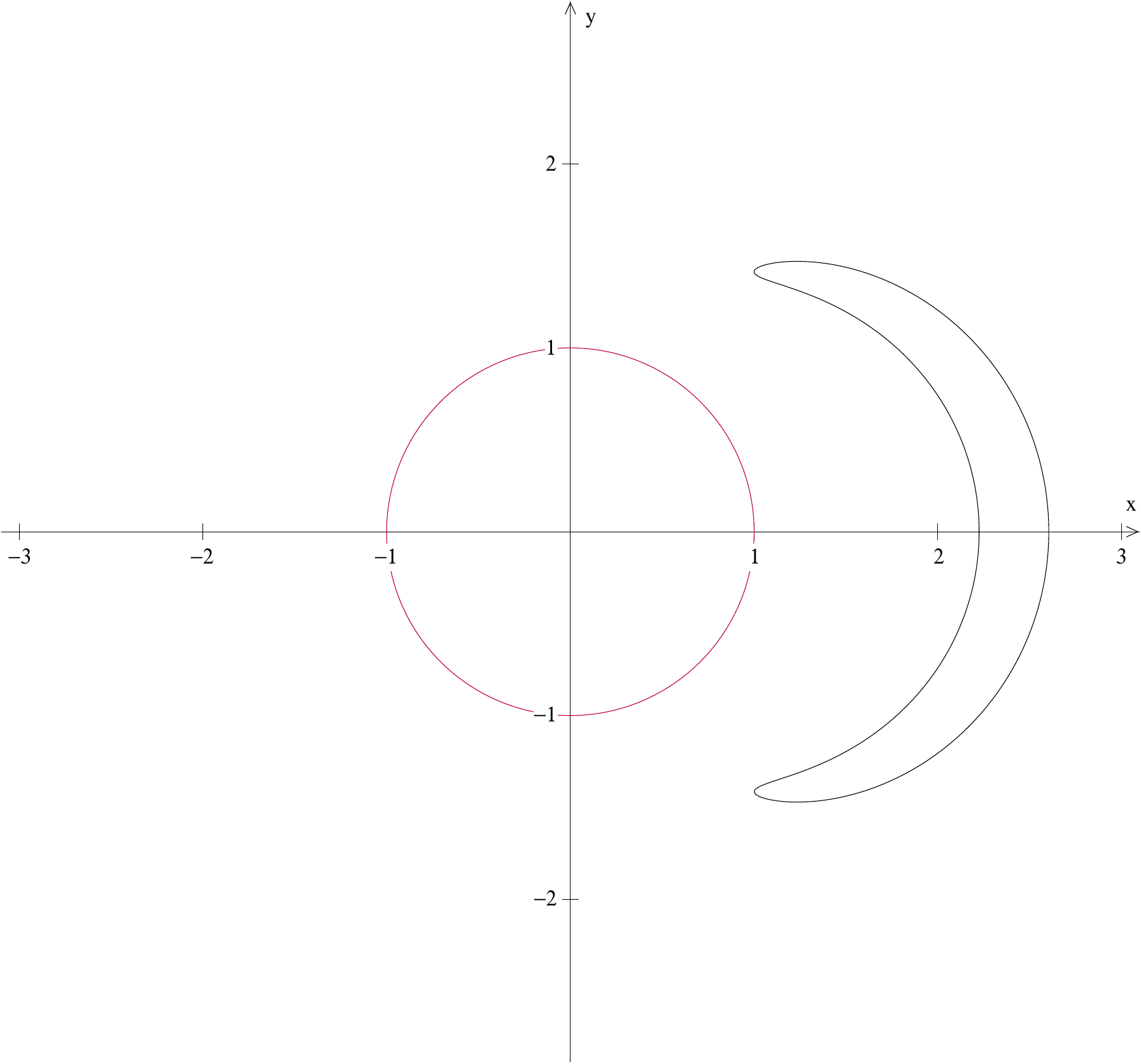}}
  \caption{Perturbations for breaking symmetry  with $k=64$ (left) and $k=10^5$ (right).} \label{fig-vz}
\end{figure}

The second issue is crucial. The ``counterexample"  (\ref{eq-vz}) has two global minimal solutions due to the symmetry
(see Fig. \ref{fig-vz}). Similar to Example 1, the canonical dual problem (\ref{eq-cdmax}) $\max \{ \Pi^d(\bsig_0,\vsig) | (\bsig_0,\vsig) \in \calS_c^+\}$
has two  stationary points on the boundary of   $\calS_c^+$ (cf.  Fig. 1(b)). Such case has been  discussed by Gao  in integer programming problem \cite{gao-jimo07}.
It was first realized that many so-called NP-hard problems in global optimization  usually have multiple global minimal solutions
and a conjecture was proposed in \cite{gao-jimo07}, i.e. a global optimization problem is NP-hard if its canonical dual has no  stationary point in $\calS_c^+$.
In order to solve such challenging problems, different perturbation methods have been suggested with successful applications in global optimization
\cite{gao-ruan-jogo10,ruan-gao-pe,r-g-j,wang-ea}, including a recent paper on solving hard-case of a trust-region subproblem \cite{chen-gao-oms}.
 For this problem, by simply using linear perturbation $\bff_k =( \frac{\sqrt{6}}{96} ,  \frac{1}{k})$ with $|k| \gg 1$,
both  global minimal solutions can be easily obtained by the canonical duality-triality theory
 \cite{mora-gao-jogo} (see Fig.  \ref{fig-vz} and Fig. 1(a)).
 Therefore,  the duality on this level is:  Voisei and Z\u{a}linescu's
 ``counterexample" does not contradict the canonical duality-triality theory even in this crucial case.

Actually, by  the general  model (\ref{eq-gs}), the nonconvex hyper-surface $h(\by) $ in this paper can be written as
$h(\by) = \WW(\DD \by) - \FF(\by)$, where the double-well  function $\WW(\DD\by)$ is objective (also isotropic), which represents the modeling with symmetry;
 while the linear term  $\FF(\by)$ is a subjective function, which breaks the symmetry and leads to a particular problem.
 By the fact that nothing in this world is  perfect, therefore, any real-world problem must have certain
 input or defects. This simple truth  lays a foundation for the perturbation method and the triality theory for solving
 challenging problems.
 However, this fact is not well-recognized in mathematical optimization and computational science\footnote{Indeed, one   authors' paper \cite{ruan-gao-pe} was first submitted to a  computational optimization journal and received   such a reviewer's comment:
 ``the authors applied a perturbation, which changed the problem
mathematically, ... and I suggest an immediate rejection."},
 it turns out that many challenges and NP-hard problems are   artificially proposed.

\subsection{Group 2: Conceptual duality}
Of four papers in this group, two  were published in pure math journals \cite{vz2,vz1}
and two were rejected by applied math  journals ({\em ZAMP} and {\em Q.J. Mech. Appl. Math}).
The paper by  Voisei and \Z \cite{vz1} challenges Gao and Strang's original work on solving the general nonconvex variational problem (\ref{eq-gs}) in finite deformation theory.
As we discussed in Section \ref{sec-cdt} that the stored energy $\WW(\beps) $ must be objective and can't be linear,
the deformation operator $\Lam$ should be geometrically admissible in order to have the canonical transformation $\WW(\beps) =\Phi(\Lam(\beps))$,
and the external energy $\FF(\bchi)$ must be linear such that $\barbchi^* = \partial \FF(\bchi)$ is the given input.
Oppositely, by listing   total six counterexamples,  Voisei and \Z  
 choose a piecewise linear function $g(u,v) = \{ u  \; (\mbox{if } v=u^2 )\; ; \;0 \mbox{ (otherwise)} \}$
as $\Phi(\bxi)$, a parametric function  $f(t) = (t, t^2)$ as the geometrically nonlinear operator $\Lam(t)$
(see Example 3.1 in \cite{vz1}), and
  quadratic functions as  $\FF(\bchi)$
(see Examples 3.2, 3.4,  3.5 and 3.6 in \cite{vz1}).
While in the rest  counterexample (Example 3.3 in \cite{vz1}),   they simply let the external energy $\FF(u) = 0$ and $\Lam(u) = u^2 -u$.

Clearly, the piecewise linear function listed by Voisei and \Z is not objective and can't be the stored energy for any real material.
Also, both $\Lam(t)$ and $\Lam(u)$ are simply not strain measures.
Such conceptual mistakes are repeatedly made in their recent papers, say in the  paper by
Strugariu,   Voisei,   and \Z
 (Example 3.3 in \cite{vz2}),   they let  $(x(t) , y(t) ) = A (t) = ( \half  t^2 , t ) $ be the geometrical mapping
 $\bxi(t) = \Lam(t)$ and, in their  notation,
$
f(x, y) = x y^3 ( x^2 + (x- y^4)^2 )^{-1}$ as the  stored energy $\Phi(\bxi)$.

For quadratic $\FF(\bchi)$, the input $\barbchi^* = \partial \FF(\bchi) $  depends linearly on the
output $\bchi$, which is called the {\em follower force}. In this case, the system is not conservative and the
traditional   variational methods do not apply. In order to study   such nonconservative minimization problems,
a so-called rate variational method and duality principle were  proposed  by Gao and Onat  \cite{gao-onat}.
While for  $\FF(\bchi) = 0$, the minimization  $\min \{\PP(\bchi) =  \WW(\DD\bchi)\}$ is  not a problem but a modelling,
which has either trivial solution $\bchi = 0$ or multiple solutions
$\bchi = constant$ due to certain symmetry of the mathematical modelling. This is a key mistake
happened very often  in global optimization,  which leads to many man-made NP-hard problems as we discussed in the
previous  subsection.

The concept of a Lagrangian was introduced by J.L. Lagrange in analytic mechanics 1788, which has a standard notation in physics as
(see \cite{land-lif})
\eb
L(\bchi) = T(  \dbchi) - \VV(\bchi),
\ee
where $\TT$ is the kinetic energy and $\VV$ is the potential energy.
By the Legendre transformation $\TT^* (\bp) = \la \dbchi, \bp \ra - \TT(\dbchi) $, the Lagrangian is also written as
\eb
L(\bchi, \bp) =  \la \dbchi, \bp \ra - \TT^*(\bp) - \VV(\bchi).
\ee
It is commonly known  that for
  problems   with  linear potential $\VV(\bchi) = \la \bchi, \barbchi^*\ra$,
  the  Lagrangian $L(\bchi) $ is convex and 
  $L(\bchi, \bp) $  is a saddle point functional which leads to a well-known  min-max  duality
  in convex systems.
But  for  problems with convex potential   $\VV(\bchi)$,
the Lagrangian  $L(\bchi) $  is a d.c. function (difference of convex functions)
and  $L(\bchi, \bp) $  is not a saddle functional any more. In this case, the Hamiltonian $H(\bchi, \bp) = \la \dbchi, \bp \ra  - L(\bchi, \bp) =  \TT^*(\bp)  +  \VV(\bchi)$ is convex. 
Therefore,   a {\em bi-duality} (i.e. the combination of the
 double-min and double-max dualities) was proposed in convex Hamilton systems (see Chapter 2 \cite{gao-dual00}).
However, in the paper by Strugariu,   Voisei,   and \Z
 \cite{vz2}, the function
 \[
 L(x,y) = \la a, x \ra \la b ,  y \ra  - \half \alp \| x\|^2 - \half \beta \| y \|^2
 \]
is defined as the ``Lagrangian", by which, they produced several ``counterexamples" for the bi-duality in convex Hamilton systems.
In this ``Lagrangian", if we consider $\VV(x) = \half \alp \| x\|^2$ as a potential energy and $\TT^*(y) =  \half \beta \| y \|^2$
as the complementary kinetic  energy, but the term $\la a, x \ra \la b ,  y \ra$ is not the bilinear form
$\la \DD x; y\ra$
required in 
  Lagrange mechanics, where $\DD$ is a differential operator  such that $\DD x $ and $y$ form a (constitutive) duality pair. This term  does not make any sense
  in Lagrangian mechanics \cite{land-lif} and duality theory \cite{eke-tem}.
  Therefore, the ``Lagrangian" used   by  Strugariu,   Voisei,   and \Z
 for producing counterexamples of the bi-duality theory  is not  the
 Lagrangian used in Gao's book \cite{gao-dual00}, i.e. the standard Lagrangian   in  classical mechanics \cite{land-lif,panza}, convex analysis \cite{eke-tem}, and modern physics \cite{davier,a-e}. Actually, the bi-duality theory    in finite dimensional space is a corollary of the so-called {\em Iso-Index Theorem} 
and the proof was given  in Gao's book 
 (see Theorem 5.3.6 and Corollary 5.3.1 \cite{gao-dual00}). 

 Papers in this group show a big gap between mathematical physics/analysis  and   optimization.
As   V.I. Arnold said \cite{arnold}: ``In the middle of the twentieth century it was attempted to divide physics and mathematics. The consequences turned out to be catastrophic."

\subsection{Group 3: Anti-Triality}  Six papers are  in this group on  the triality theory. By listing simple
  counterexamples (cf. e.g. \cite{vz-jogo}),  Voisei  and \Z claimed:
``a correction of this   theory is impossible without falling into trivia"\footnote{This sentence is deleted
by Voisei  and \Z in their revision of \cite{vz-jogo} after they were informed by referees that their counterexamples are not new
and the triality theory has been proved.}.
However, even some of these counterexamples are correct, they are   not new. This type of counterexamples  was  first discovered  by Gao in  2003
 \cite{gao-amma03,gao-opt03}, i.e. the double-min duality holds under certain additional constraints  (see  Remark on page 288 \cite{gao-amma03} and
 Remark 1 on page 481 \cite{gao-opt03}).
 But neither \cite{gao-amma03} nor \cite{gao-opt03} was cited by Voisei  and \Z  in their papers.

 As  mentioned in Section \ref{sec-tri},  the triality was  proposed originally from  post-buckling analysis   \cite{gao-amr97}
 in ``either-or" format
 since the double-max duality is always true but the double-min duality was proved only in   one-dimensional nonconvex analysis \cite{gao-dual00}.
 Recently, this double-min duality has been proved  first for polynomial optimization    \cite{gao-wu-jimo,mora-gao-naco,mora-gao-memo},
  and then for general global optimization problems \cite{chen-gao-jogo,gao-wu-jogo}.
  The �certain additional constraints� are simply the dimensions of the primal problem and its canonical dual should be the same in order to have strong double-min duality. Otherwise, this double-min duality holds weakly in   subspaces with elegant symmetrical forms. Therefore, the triality theory
  now has been proved in  global optimization, which should play important roles for  solving NP-hard problems in complex systems.




\section{Concluding  Remarks and Open Problems}
In this article  we have discussed the existing gaps between nonconvex analysis/mechanics and global optimization.
Common misunderstandings and confusions on some basic concepts have been addressed and clarified, including
the objectivity, nonlinearity, and Lagrangian.
By the fact that the canonical duality is a fundamental law in nature, the canonical duality-triality theory
is indeed  powerful for unified understanding complicated phenomena and solving challenging problems.
So far, this theory can be summarized for  having    the following functions:

\begin{verse}
1.    To correctly model complex phenomena in multi-scale systems within a unified framework \cite{gao-dual00,gao-amma03,gao-yu}.\\

2.    To solve a large class of nonconvex/nonsmooth/discrete  global optimization  problems  for obtaining
 both global and local optimal solutions. \\

3.    To reformulate certain nonlinear partial differential equations in algebraic forms with possibility to obtain all
 possible analytical solutions  \cite{gao-na00,gao-anti,gao-ogden-zamp,gao-ogden-qjmam}.\\

4.    To understand and identify certain NP-hard problems, i.e., the general  global optimization problems are not NP-hard if they can be solved by the canonical duality-triality theory \cite{gao-jimo07,gao-ruan-jogo10,ruan-gao-pe}.\\

5.    To understand and solve nonlinear (chaotic) dynamic systems by obtaining global stable solutions \cite{ruan-gao-ima,li-zhou-gao}.\\

6.      To check and verify correctness of existing  modelling and theories.

\end{verse}

There are still  many open problems existing in the canonical duality-triality theory.
Here we list a few of them.
\begin{verse}
1. Sufficient condition for the existence of the canonical dual solutions on $\calS^+_c$.\\
2. NP-Harness conjecture: A global optimization problem is NP-hard if its canonical dual $\Pi^d(\bxi^*)$ has no stationary point on the closed domain
$\bar{\calS}^+_c = \{ \bxi^* \in\calS_a| \; \bG(\bxi^*  \succeq 0 \}$.\\
3.Extremality conditions for stationary points of $\Pi^d(\bxi^*)$ on the domain such that $\bG(\bxi^*)$ is in-definite in order to identify all local extrema. \\
4.  Bi-duality and triality theory for $d$-dimensional ($d> 1$)  nonconvex analysis problems.
\end{verse}

The following research topics are challenging:

1.
Canonical duality-triality theory for solving bi-level optimization problems.

2. Using least-squares method and canonical duality theory for solving 3-dimensional chaotic dynamical problems, such as
Lorenz system and Navier-Stokes equation, etc.

3. Perturbation methods for solving NP-hard integer programming problems, such as quadratic Knapsack problem, TSP, and mixed integer nonlinear programming problems.

4. Unilateral post-buckling problem  of the Gao nonlinear  beam
\eb
\min_{\chi \in \calX_a} \left\{ \Pi(\chi) = \int_0^L \left[ \half EI  \chi_{xx}^2 +  \frac{1}{12} \alp E \chi_{x}^4 - \half \lam E \chi_{x}^2 - f \chi \right] \dx
\;\;  | \;\;
\chi(x) \ge 0 \right\} .
\ee
Due to the axial compressive load $\lam > 0$, the
downward lateral load $f(x) $ and the  unilateral constraint $\chi(x) \ge 0 \;\; \forall x\in [0,L]$,
the solution of this nonconvex variational problem is a local minimizer of $\Pi(\chi)$ which can be obtained numerically
by the canonical dual finite element methods \cite{cai-gao-qin,santos-gao} if $\lam$ and $f$ are not big enough such that
$\chi(x) > 0 \;\;\forall x\in [0,L]$.
However, if the buckling state  $\chi(x) = 0 $ happens  at any $x\in [0, L]$, the problem
could be NP-hard.
The open problems include:

1) under what conditions for the external loads $\lam > 0 $ and $f(x)$, the problem
has a solution $\chi(x) > 0 \;\;\forall x \in [0,L]$?

2) how to solve the unilateral buckling problem when $\chi(x) =  0  $ holds for certain $x \in  [0,L]$? \\

\noindent{\bf Acknowledgement}:
This paper is based on  a series of plenary lectures presented   at
  international conferences of mathematics,  mechanics and global optimization during 2012-2014.
  Invitations from organizers are sincerely acknowledged.
The research has been supported by  a grant (AFOSR FA9550-10-1-0487)
from the US Air Force Office of Scientific Research. Dr. Ning Ruan was
supported by a funding from the Australian Government
under the Collaborative Research Networks (CRN) program.
The authors sincerely thank Professor D. Steigmann at University of Berkeley for his invitation of  editing the special issue of
the journal {\em Math. Mech. Solids}
on the canonical duality theory.


\end{document}

%% file: command.tex
\newcommand{\sta}{{\rm sta}}

\newcommand{ \PP}{{P}}

\newcommand{\vsig}{\varsigma}

\newcommand{\real}{{\mathbb R}} 

\newcommand{\half}{\frac{1}{2}}

\newcommand{\bsig}{\mbox{\boldmath$\sigma$}}
\newcommand{\bvsig}{\mbox{\boldmath$\varsigma$}}
\newcommand{\bmu}{\mbox{\boldmath$\mu$}}

\newcommand{\beps}{\mbox{\boldmath$\epsilon$}}

\newcommand{\btau}{{\mbox{\boldmath$\tau$}}}

\newcommand{\blam}{{\mbox{\boldmath$\lambda$}}}
\newcommand{\bLam}{{\mbox{\boldmath$\Lambda$}}}
\newcommand{\barbtau}{\bar{\mbox{\boldmath$\tau$}}}
\newcommand{\bchi}{{\mbox{\boldmath$\chi$}}}

\newcommand{\barbchi}{\bar{\mbox{\boldmath$\chi$}}}
\newcommand{\bxi}{{\mbox{\boldmath$\xi$}}}

\newcommand{\barbxi}{\bar{\mbox{\boldmath$\xi$}}}

\newcommand{\la}{\langle}
\newcommand{\ra}{\rangle}

\newcommand{\bp}{{{\bf p}}}
\newcommand{\Oo}{\Omega}

\newcommand{\Gt}{{\Gamma_t}}
\newcommand{\Gu}{{\Gamma_u}}
\newcommand{\bE}{{\bf E}}
\newcommand{\bQ}{{\bf Q}}

\newcommand{\eba}{\begin{array}}
\newcommand{\eea}{\end{array}}
\newcommand{\ebe}{\begin{eqnarray}}
\newcommand{\eee}{\end{eqnarray}}
\newcommand{\eb}{\begin{equation}}
\newcommand{\ee}{\end{equation}}
\newcommand{\bT}{{\bf T}}

\newcommand{\calW}{{\cal{W}}}
\newcommand{\calP}{{\cal{P}}}

\newcommand{\bg}{{\bf g}}
\newcommand{\bC}{{\bf C}}
\newcommand{\bG}{{\bf G}}
\newcommand{\bH}{{\bf H}}
\newcommand{\bn}{{\bf n}}
\newcommand{\bt}{{\bf t}}
\newcommand{\bff}{{\bf f}}

\newcommand{\bR}{{\bf R}}
\newcommand{\bq}{{\bf q}}
\newcommand{\bv}{{\bf v}}
\newcommand{\bw}{{\bf w}}

\newcommand{\bx}{{\bf x}}
\newcommand{\by}{{\bf y}}

\newcommand{\bX}{{\bf X}}

\newcommand{\bN}{{\bf N}}
\newcommand{\bF}{{\bf F}}
\newcommand{\bA}{{\bf A}}
\newcommand{\bB}{{\bf B}}
\newcommand{\bI}{{\bf I}}

\newcommand{\calS}{{\cal S}}

\newcommand{\calC}{{\cal C}}
\newcommand{\calT}{{\cal T}}

\newcommand{\calU}{{\cal U}}
\newcommand{\calE}{{\cal E}}

\newcommand{\calV}{{\cal V}}

\newcommand{\calY}{{\cal Y}}
\newcommand{\calX}{{\cal X}}

\newcommand{\barbS}{\bar{\bf S}}

\newcommand{\barbp}{\bar{\bf p}}
\newcommand{\barbq}{\bar{\bf q}}

\newcommand{\bb}{{{\bf b}}}

\newcommand{\bS}{{\bf S}}

\newcommand{\dt}{{\,\mbox{d}t}}

\newcommand{\be}{{\bf e}}

\newcommand{\dx}{\mbox{ d}x}

\newcommand{\dO}{\,\mbox{d}\Oo}

\newcommand{\dG}{\,\mbox{d} \Gamma}

\newcommand{ \grad }{{\mbox{grad}}}
\newcommand{\alp}{{\alpha}}

\newcommand{ \eps}{{\epsilon}}

\newcommand{ \sig}{{\sigma}}
\newcommand{ \Lam}{{\Lambda}}

\newcommand{ \lam}{{\lambda}}
\newcommand{ \xx}{{\bf x}}

\newcommand{\bc}{{\bf c}}
\newcommand{\bD}{{\bf D}}

\newcommand{\tr}{{\mbox{tr}}}

\newcommand{\Diag}{{\mbox{Diag }}}

\newcommand{\curl}{{\mbox{curl}}}

\newtheorem{remark}{Remark}

\newtheorem{thm}{Theorem}

\newtheorem{definition}{Definition}

\newtheorem{example}{Example}